\documentclass[a4paper,onecolumn,superscriptaddress,11pt,accepted=2019-09-07]{quantumarticle}
\pdfoutput=1
\usepackage{qcircuit}
\usepackage[dvips]{graphicx}
\usepackage{amsmath,amssymb,amsthm,mathrsfs,amsfonts,dsfont}
\usepackage{subfigure, epsfig}
\usepackage{braket}
\usepackage{bm}
\usepackage{enumerate}
\usepackage{color}
\usepackage{multirow}
\usepackage{booktabs}
\usepackage{arydshln}
\usepackage{colortbl}

\usepackage[table]{xcolor}
\usepackage{textcomp}
\definecolor{quantumviolet}{HTML}{53257F}
\usepackage{hyperref}
\hypersetup{citecolor=quantumviolet}
\hypersetup{colorlinks=true}
\hypersetup{linkcolor=quantumviolet}
\hypersetup{urlcolor=quantumviolet}

\usepackage{multirow}
\usepackage[utf8]{inputenc}  
\usepackage[british]{babel}  
\usepackage{graphicx} 
\usepackage[babel]{microtype}  
\usepackage{amsmath,amssymb,bm,amsfonts,mathrsfs,bbm} 
\usepackage{xspace}  
\usepackage{pgfplots}  %
\usepackage{colortbl}
\usepackage{braket}

\usepackage{graphics}

\usepackage[numbers,sort&compress]{natbib}

\newcommand{\tr}{\mathrm{Tr}}

\newcommand{\phitau}{\phi(\vec{\theta}(\tau))}
\newcommand{\phit}{\phi(\vec{\theta}(t))}
\newcommand{\rhot}{\rho(\vec{\theta}(t))}
\newcommand{\ml}{\mathcal{L}(\rho) }

\newcommand{\HC}{{\rm h.c.}}
\newcommand{\LL}{\mathcal{L}}
\newcommand{\RR}{\mathcal{R}}

\newcommand{\Tr}{\mathrm{Tr}}

\newcommand{\yx}[1]{{#1}}
\newcommand{\yxx}[1]{{#1}}

\begin{document}

\title{Theory of variational quantum simulation}

\author{Xiao Yuan}
\affiliation{Department of Materials, University of Oxford, Parks Road, Oxford OX1 3PH, United Kingdom}

\orcid{0000-0003-0205-6545}

\author{Suguru Endo}
\affiliation{Department of Materials, University of Oxford, Parks Road, Oxford OX1 3PH, United Kingdom}

\author{Qi Zhao}
\affiliation{Center for Quantum Information, Institute for Interdisciplinary Information Sciences, Tsinghua University, Beijing 100084, China}

\author{Ying Li}
\affiliation{Graduate School of China Academy of Engineering Physics, Beijing 100193, China}

\author{Simon C.~Benjamin}
\affiliation{Department of Materials, University of Oxford, Parks Road, Oxford OX1 3PH, United Kingdom}
\orcid{0000-0002-7766-5348}

\maketitle

\begin{abstract}
  The variational method is a versatile tool for classical simulation of a variety of quantum systems. Great efforts have recently been devoted to its extension to quantum computing for efficiently solving static many-body problems and simulating real and imaginary time dynamics. In this work, we first review the conventional variational principles, including the Rayleigh-Ritz method for solving static problems, and the Dirac and Frenkel variational principle, the McLachlan's variational principle, and the time-dependent variational principle, for simulating real time dynamics. We focus on the simulation of dynamics and discuss the connections of the three variational principles. Previous works mainly focus on the unitary evolution of pure states. In this work, we introduce variational quantum simulation of mixed states under general stochastic evolution. We show how the results can be reduced to the pure state case with a correction term that takes accounts of global phase alignment. For variational simulation of imaginary time evolution, we also extend it to the mixed state scenario and discuss variational Gibbs state preparation. We further elaborate on the design of ansatz that is compatible with post-selection measurement and the implementation of the generalised variational algorithms with quantum circuits. Our work completes the theory of variational quantum simulation of general real and imaginary time evolution and it is applicable to near-term quantum hardware.
\end{abstract}

\tableofcontents

\section{Introduction}
Variational simulation is a widely used technique in many-body physics \cite{BALIAN198829, PhysRevA.56.1424, RevModPhys.71.463, PhysRevLett.107.070601, Verstraete04, SHI2018245, vanderstraeten2018tangent} and chemistry \cite{RevModPhys.72.655,szabo2012modern,helgaker2014molecular}.
As the full Hilbert space of a quantum system grows exponentially with respect to the system size, direct classical simulation of a large many-body state is generally impossible.
The variational method avoids the exponential space problem by considering trial states from a physically motivated small subset of the exponentially large Hilbert space \cite{verstraete2008matrix,ashida2018variational}. Variational classical simulation (VCS) works for both static problems in finding the ground state and ground state energy, and dynamical problems in simulating the real and imaginary time evolution of quantum states \cite{JACKIW1979158,LEHTOVAARA2007148,Verstraete04,Kramer08,Christina10,PhysRevLett.107.070601,SHI2018245,ashida2018variational}. In modern science, the variational method has become a versatile tool for simulating various problems when the target system state can be well modelled classically. 

Nevertheless, there also exist many problems that may not be solved classically even with the classical variational method \cite{harrow2017quantum,boixo2018characterizing,neill2018blueprint}. This is because there exist highly entangled many-body states that may not be efficiently represented via any classical method. This problem motivates the development of quantum simulation with universal quantum computers \cite{Feynman1982}, because quantum systems can be efficiently encoded or represented by another quantum system and real time evolution of the Schr\"odinger equation can be efficiently realised via a unitary quantum circuit \cite{Lloyd1073,Abrams97}. However, implementing a universal quantum computer requires the coherent and accurate control of millions of qubits \cite{PhysRevA.95.032338, campbell2017roads, Reiher201619152}; While the state-of-the-art quantum hardwares can only accurately control tens of qubits \cite{wooten2018cloud,neill2018blueprint}. Although this number may be extended by one or two orders in the near future, realising a universal fault tolerant quantum computer remains a challenging task.

With noisy intermediate-scale quantum hardware~\cite{preskill2018quantum}, quantum advantages may still be achieved in many tasks with recently proposed hybrid or, more specifically, variational quantum simulation (VQS) methods \cite{farhi2014quantum,peruzzo2014variational,wang2015quantum,PRXH2,PhysRevA.95.020501,VQETheoryNJP,PhysRevLett.118.100503,PhysRevX.8.011021,Santagatieaap9646,kandala2017hardware,kandala2018extending,PhysRevX.8.031022,kokail2018self, Li2017,nakanishi2018subspace}. By dividing the problem into two levels, VQS methods only use the quantum processor to solve the core and classically intractable problem and leave the relatively easy task to a classical computer. 
For solving static problems, the Rayleigh-Ritz variational method can be naturally generalised to the quantum regime, such as the variational quantum eigensolver (VQE)~\cite{peruzzo2014variational,wang2015quantum,PRXH2,PhysRevA.95.020501,VQETheoryNJP,PhysRevLett.118.100503,PhysRevX.8.011021,Santagatieaap9646,kandala2017hardware,kandala2018extending,PhysRevX.8.031022,kokail2018self} and the quantum approximate optimisation algorithm (QAOA)~\cite{farhi2014quantum}.
Considering parameterised trial states prepared by quantum circuits \cite{PhysRevLett.106.170501,RevModPhys.86.153,whaley2014quantum}, the target static problem is mapped to a cost function of the measurement outcomes of the quantum state, which is further optimised via a classical algorithm. 
As the state preparation and measurement procedures may be hard to simulate by a classical computer \cite{harrow2017quantum,boixo2018characterizing,neill2018blueprint}, hybrid or variational quantum simulation can have an intrinsic quantum advantage over VCS and it has been proposed as the standard tool for studying quantum computational chemistry \cite{2018arXiv180810402M,cao2018quantum}.  

A broader and harder challenge is the simulation of dynamics. 
Classically, there are three variational principles---the Dirac and Frenkel variational principle~\cite{dirac1930note,frenkel1934wave}, the McLachlan variational principle~\cite{McLachlan}, and the time-dependent variational method principle~\cite{kramer1981geometry,broeckhove1988equivalence}. 
\yx{These variational principles have been extensively studied in VCS for example on Bose-Einstein condensation~\cite{PhysRevA.56.1424} and with matrix-product states in the so called tangent-bundle formalism~\cite{PhysRevB.88.075133}.}
Recently, the variational methods have been generalised to simulating the real~\cite{Li2017,heya2019subspace} and imaginary~\cite{mcardle2018variational,endo2018discovering} time evolution of pure quantum systems, and been experimentally implemented with superconducting qubits~\cite{2019arXiv190503150C}. 
By encoding quantum states with a parameterised circuit, the evolution of the state can be mapped to the evolution of the parameters controlling the circuit.  
However, different variational principles can lead to different evolution equations of parameters.  It has not been clear which variational principle or which evolution equation should be used in practice. Furthermore, previous works only focus on the evolution of pure states of closed systems, rather than the more general case of the stochastic evolution of mixed states of open systems. While practical quantum systems always suffer from unavoidable interaction with the environment, it makes almost every quantum system open. Therefore, simulating the evolution of open quantum systems can be useful for studying realistic quantum systems.

In this work, we complete the VQS theory by studying the equivalence and difference of the variational principles and extending the principles to the general stochastic evolution of mixed states.
We first review the theory of variational simulation in Sec.~\ref{Sec:preliminary}.
Then we study the equivalence and difference of the variational principles and the derived evolution equations in Sec.~\ref{Sec:equivalence}. \yx{We summarise the results in Table~\ref{tab:results} under various conditions.}
In particular, we consider general real time stochastic evolution of mixed states in Sec.~\ref{Sec:real}. As an example, we re-derive the evolution equation of parameters for unitary evolution of pure states and find a correction term which applies to previous results and handles the global phase alignment. 
After applying all the three variational principles to the general mixed state case, we find that McLachlan's variational principle is the most appropriate variational principle that leads to a consistent theory of VQS. 
In Sec.~\ref{Sec:imaginary}, we study variational simulation of imaginary time evolution of mixed states and discuss variational Gibbs state preparation.
We further elaborate on the implementation with quantum circuits in Sec.~\ref{Sec:implementation}.
We conclude our work and discuss future directions in Sec.~\ref{Sec:Discussion}.

\section{Preliminary}\label{Sec:preliminary}
In this section, we introduce the background of variational classical and quantum simulation. We first review the Rayleigh-Ritz principle for solving static problems and the Dirac and Frenkel variational principle \cite{kramer1981geometry,broeckhove1988equivalence}, the McLachlan variational principle \cite{McLachlan}, and the time-dependent variational principle \cite{kramer1981geometry,broeckhove1988equivalence} for simulating dynamics. We also review the implementation of the simulation with quantum circuits.
\yx{It is worth noting that the variational principles work similarly for both variational classical and quantum simulation.  The main differences between variational classical and quantum simulation are about how the trial state or the ansatz is implemented and how the evolution equation of the parameters is determined, as we shortly discuss. Variational classical simulation has been extensively studied for different tasks~\cite{JACKIW1979158,LEHTOVAARA2007148,Verstraete04,Kramer08,Christina10,PhysRevLett.107.070601,SHI2018245,ashida2018variational}. This section will not review those results. Instead, we aim to have a self-consistent introduction of different variational principles with a focus on variational quantum simulation.}

\subsection{Static problem: Rayleigh-Ritz method}
The Rayleigh-Ritz method is designed to solve static problems of finding the ground state energy $E_0$ of a given Hamiltonian $H=\sum_jh_j\sigma_j$. Here, we assume that $H$ is a linear combination of tensor products of local operators $\sigma_j$ with coefficients $h_j$.
The ground state energy is the solution of the  problem,
\begin{equation}
	E_0 = \min_{\ket{\psi}}\frac{\bra{\psi}H\ket{\psi}}{\braket{\psi|\psi}},
\end{equation} 
where the minimisation is over states from the whole Hilbert space. Because the size of the Hilbert space grows exponentially to the system size, it is in general computationally hard to brutal forcely search the whole Hilbert space. Instead, the Rayleigh-Ritz method only searches states from a subset of the whole Hilbert space to find an approximate solution. 

For example, classically, one can consider trial states that are linear combinations of a small number of basis states $\{\ket{\phi_i}\}$,
\begin{equation}
	\ket{\phi(\vec{a})} = \sum_ia_i\ket{\phi_i},
\end{equation}
with $\vec{a} = (a_1, a_2, \dots)$. Suppose the ground state can be approximated by the 
trial state $\ket{\phi(\vec{a})}$ with a proper vector of coefficients $\vec{a}$, then the ground state energy can be approximated or upper bounded by
\begin{equation}
\begin{aligned}
	E_0 \le E_0^{est}=& \min_{\vec{a}}\frac{\bra{\phi(\vec{a})}H\ket{\phi(\vec{a})}}{\braket{\phi(\vec{a})|\phi(\vec{a})}}
	=\min_{\vec{a}}\frac{\sum_{i,j}a_i^*a_j\bra{\phi_i}H\ket{\phi_j}}{\sum_{i,j}a_i^*a_j\braket{\phi_i|\phi_j}}.
\end{aligned}
\end{equation}
\yx{Suppose each term $\bra{\phi_i}H\ket{\phi_j}$ can be efficiently obtained}, the minimisation problem can be solved by calculating the partial derivative of $E_0^{est}$ over $a^*_i$, and is equivalent to the minimal solution to
\begin{equation}
	\det(H-\lambda S)=0,
\end{equation}
where $H$ and $S$ are the matrix of $\bra{\phi_i}H\ket{\phi_j}$ and $\braket{\phi_i|\phi_j}$, respectively.
We refer to the trial state also as the ansatz. 
In practice, different ans\"atze have been invented in condensed matter physics and computational chemistry \cite{verstraete2008matrix,RevModPhys.86.153,SHI2018245,whaley2014quantum,2018arXiv180810402M}.
When the minimisation cannot be solved analytically, one can also numerically solve it with classical optimisation algorithms. \yx{Such a variational method with classical ansatz is called VCS. }

With quantum computers or in VQS, the trial state $\ket{\phi(\vec{\theta})}$ can be prepared by applying a sequence of parameterised gates $R_{k}(\theta_k)$ to an initial state $\ket{\bar{0}}$ as 
$$\ket{\phi(\vec{\theta})}  = R_{N}(\theta_N)\dots R_{k}(\theta_k)\dots R_{1}(\theta_1)\ket{\bar{0}}.$$ Here, $R_{k}(\theta_k)$ is the $k^{\textrm{th}}$ gate controlled by the real parameter $\theta_k$ and $\vec{\theta}= (\theta_1, \theta_2,\dots, \theta_N)$. The ansatz is automatically normalised and the average energy $\braket{\phi(\vec{\theta})|H|\phi(\vec{\theta})}$ can be obtained by measuring each term $\braket{\phi(\vec{\theta})|h_i|\phi(\vec{\theta})}$ and linearly combining the measurement outcomes. 
In Sec.~\ref{Sec:implementation}, we also introduce other possible ways of realising the trial states.
To approximate the ground state, we can minimise $\braket{\phi(\vec{\theta})|H|\phi(\vec{\theta})}$ over the parameter space with a classical optimisation algorithm. For instance, by calculating the gradient of $\braket{\phi(\vec{\theta})|H|\phi(\vec{\theta})}$, a local minimum of $\braket{\phi(\vec{\theta})|H|\phi(\vec{\theta})}$ can be found via the gradient descent algorithm. 
Such a hybrid algorithm for solving the eigenvalue of the Hamiltonian is called variational quantum eigensolver (VQE)
\cite{peruzzo2014variational,wang2015quantum,PRXH2,PhysRevA.95.020501,VQETheoryNJP,PhysRevLett.118.100503,PhysRevX.8.011021,Santagatieaap9646,kandala2017hardware,kandala2018extending,PhysRevX.8.031022,kokail2018self}.
\yx{Note that parameters can be complex for classical simulation but it suffices for them to be real for quantum simulation. This is because without loss of generality the trial state can be prepared by a quantum circuit by applying unitary gates $R_{k}(\theta_k)$ that is in the form of $e^{i\theta_k\sigma}$ with Hermitian operator $\sigma$ and real parameter $\theta_k$.}

\subsection{Dynamics of closed systems: three variational principles}
Now we introduce three variational principles for simulating real time dynamics of closed systems with pure quantum states~\cite{PhysRevLett.107.070601,Li2017}. We only review the results here and leave the detailed derivation in Appendix~A.
To simulate Schr\"odinger's equation
\begin{equation}\label{Schrodinger}
	\frac{d \ket{\psi(t)}}{d t} = -iH\ket{\psi(t)},
\end{equation}
we consider a parametrised trial state $\ket{\phi(\vec{\theta}(t))}$ with time dependent parameters. \yx{Again, the parameters can be complex for VCS and are assumed to be real for VQS.}
Suppose the state at time $t$ is represented by the trial state $\ket{\psi(t)}=\ket{\phi(\vec{\theta}(t))}$ with parameters $\vec{\theta}(t)$, then we want to approximate the state $\ket{\psi(t+\delta t)}$ at time $t+\delta t$ by $\ket{\phi(\vec{\theta}(t+\delta t))}$. \yx{Note that the state is evolved from $\ket{\psi(t)}$ to  $\ket{\psi(t+\delta t)}$ according to Schr\"odinger's equation as
$$\ket{\psi(t+\delta t)}\approx\ket{\phi(\vec{\theta}(t))} - i\delta tH\ket{\phi(\vec{\theta}(t))}.$$
Such a state 
may be impossible to be represented by the trial state with any parameters. 
A variational method is used to project the state $\ket{\psi(t+\delta t)}$ back to the trial state manifold or find the best solution $\vec{\theta}(t+\delta t)$ that approximates 
\begin{equation}\label{Eq:target}
	\ket{\phi(\vec{\theta}(t+\delta t))} \approx \ket{\phi(\vec{\theta}(t))} - i\delta tH\ket{\phi(\vec{\theta}(t))}.
\end{equation}
Note that 
$$\ket{\phi(\vec{\theta}(t+\delta t))}\approx \ket{\phi(\vec{\theta}(t))}+\sum_j\frac{\partial \ket{\phi(\vec{\theta}(t))}}{\partial \theta_j}\delta{\theta}_j,$$ 
the target problem reduces to approximate
\begin{equation}\label{Eq:target2}
	\sum_j\frac{\partial \ket{\phi(\vec{\theta}(t))}}{\partial \theta_j}\delta{\theta}_j \approx  - i\delta tH\ket{\phi(\vec{\theta}(t))}.
\end{equation}
Here, the L.H.S. and R.H.S. denote the change of the trial state by varying the parameters and the evolution according to the Schr\"odinger equation, respectively. 
Under different variational principles, we aim to minimise the difference so that the evolution of the state $\ket{\psi}$ under the Schr\"odinger equation can be mapped onto the trial state manifold as the evolution of parameters. 
}

\subsubsection{The Dirac and Frenkel variational principle}
In Eq.~\eqref{Eq:target2}, the L.H.S is in the tangent subspace $\{\frac{\partial \ket{\phi(\vec{\theta}(t))}}{\partial \theta_j}\}$, but the R.H.S. $-iH\ket{\phi(\vec{\theta}(t))}$ may not be in this subspace.
So the Dirac and Frenkel variational principle \cite{kramer1981geometry,broeckhove1988equivalence} directly projects the equation onto the subspace, which can be equivalently expressed by
\begin{equation}
	\left\langle{ \delta\phit\bigg|\left(\frac{d}{d t}+iH\right)\bigg|\phi(\vec{\theta}(t))}\right\rangle = 0,
\end{equation}
with $\bra{\delta\phit} = \sum_i\frac{\partial \bra{\phi(\vec{\theta}(t))}}{\partial \theta_i}\delta\theta_i$.
The evolution of parameters can be solved to be
\begin{equation}\label{Eq:pureUnitaryComplex}
	\sum_j A_{i,j}\dot{\theta}_j = -iC_i,
\end{equation}
where $A$ and $C$ are
\begin{equation}\label{Eq:defAandC}
	\begin{aligned}
		A_{i,j} &= \frac{\partial \bra{\phi(\vec{\theta}(t))}}{\partial \theta_i}\frac{\partial \ket{\phi(\vec{\theta}(t))}}{\partial \theta_j},\\
		C_i &= \frac{\partial \bra{\phi(\vec{\theta}(t))}}{\partial \theta_i}H\ket{\phi(\vec{\theta}(t))}.
	\end{aligned}
\end{equation}
In general, $A$ and $C$ are complex even if the parameters are real, which leads to a complex solution  $\dot{\theta}_j$. A complex solution is not problematic in classical simulation as parameters are complex. However, parameters in quantum simulation are generally real, a complex solution thus leads to a state outside of the ansatz space. 
Naively, one can take the real or imaginary part of Eq.~\eqref{Eq:pureUnitaryComplex} to make the solution real. 
Although this seems artificial, we show that the corresponding equations can be obtained from the McLachlan's variational principle and the time-dependent variational principle, respectively.

\subsubsection{McLachlan's variational principle}
The McLachlan's variational principle~\cite{McLachlan} is to minimise the distance between the L.H.S and R.H.S of Eq.~\eqref{Eq:target2}, which can be equivalently expressed as
 \begin{equation}
 	\delta \|({d}/{d t} +iH)\ket{\phit}\|=0.
 \end{equation}
\yxx{Here $\|\ket{\psi}\| = \sqrt{\braket{\psi|\psi}}$ is the norm of quantum states $\ket{\psi}$.} 
When the parameters $\theta$ are complex, the solution gives the same evolution of Eq.~\eqref{Eq:pureUnitaryComplex} under the Dirac and Frenkel variational principle, \yx{indicating the equivalence of these two variational principles}.
However, assuming $\theta$ to be real, the equation becomes
\begin{equation}\label{Eq:pureRealSolution}
	\sum_j A^R_{i,j}\dot{\theta}_j = C_i^I,
\end{equation}
which can be obtained by taking the real part of Eq.~\eqref{Eq:pureUnitaryComplex}. Here $A^R_{i,j}$ is the real part of $A_{i,j}$ and $C_i^I$ is the imaginary part of $C_i$. As shown in Sec.~\ref{Sec:equivalence}, this equation is equivalent to Eq.~\eqref{Eq:pureUnitaryComplex} when the trial state $\ket{\phit}$ is powerful enough to represent the target state $\ket{\psi(t)}$. 
The advantage of this equation is that it always gives a real solution of $\dot{\theta}$, which makes it consistent with VQS with real parameters. 
Furthermore, we will show shortly that McLachlan's variational principle is the most consistent variational principle that works similarly to general stochastic evolution. 

\yx{However, it is worth noting that the evolution according to Eq.~\eqref{Eq:pureRealSolution} is not the ultimate one owing to a time-dependent global phase mismatch between the target state $\ket{\psi(t)}$ and the trial state $\ket{\phit}$. 
That is, even if $\ket{\phit}$ can represent $\ket{\psi(t)}$ up to a global phase, Eq.~\eqref{Eq:pureRealSolution} may still be incorrect to be the evolution equation of the parameters. The main reason for this seemly counter-intuitive phenomenon stems from the fact that even if $\ket{\psi(t)}$ and $\ket{\phit}$ are equivalent up to a time-dependent global phase,  their derivatives $d\ket{\psi(t)}/dt$ and $d\ket{\phit}/dt$ can be very different.}
\yx{This problem can be either addressed by introducing an actual time-dependent phase gate with an additional parameter as $e^{i\theta_{0}(t)}\ket{\phit}$. Alternatively, we show that this is not necessary as the problem can be resolved with a modification of Eq.~\eqref{Eq:pureRealSolution}. Suppose the state at time $t$ is represented by the trial state, 
$\ket{\psi(t)}=e^{i\theta_{0}(t)}\ket{\phit}$, up to a global phase $\theta_{0}(t)$. Although global phase is physically irrelevant, it plays an important role in the variational method. Considering Eq.~\eqref{Eq:target2}, the L.H.S. becomes
$$
	e^{i\theta_{0}(t)}\sum_j\frac{\partial \ket{\phi(\vec{\theta}(t))}}{\partial \theta_j}\delta{\theta}_j + ie^{i\theta_{0}(t)}\delta\theta_0\ket{\phit}
$$
and the R.H.S. is $-i\delta tHe^{i\theta_{0}(t)}\ket{\phit}$. According to McLachlan's variational principle, the minimisation between the difference of them gives a similar evolution equation of the parameters $\vec{\theta}$,
\begin{equation}
	\sum_j M_{i,j} \dot\theta_j = V_i,
\end{equation}
with
\begin{equation}\label{}
	\begin{aligned}
		M_{i,j} &= A_{i,j}^R+ \frac{\partial \bra{\phi(\vec{\theta}(t))}}{\partial \theta_i}\ket{\phit}\frac{\partial \bra{\phi(\vec{\theta}(t))}}{\partial \theta_j}\ket {\phit},\\ 
		V_i &= C_i^I+i\frac{\partial \bra{\phi(\vec{\theta}(t))}}{\partial \theta_i}\ket{\phit}\bra{\phit}H\ket{\phi(\vec{\theta}(t))}.
	\end{aligned} 
\end{equation}
Therefore, as long as we measure the correction terms, we can automatically resolve the global phase problem without actually introducing the global phase. 
As shown in Sec.~\ref{Sec:real}, those correction terms can be also obtained by considering the McLachlan's variational principle for mixed states, which automatically handles the global phase problem. 
}
%

A further point of interest concerns our ability to track the accuracy of the simulation.
In practice, after solving the derivatives of $\dot\theta$ in Eq.~\eqref{Eq:pureRealSolution}, we can also get the distance between the true evolution and the evolution of the trial state, 
\begin{equation}
	\|({d}/{d t} +iH)\ket{\phit}\|^2 = \sum_{i,j}A^R_{i,j}\dot\theta_i\dot\theta_j - 2\sum_i C^I_i\dot\theta_i + \braket{H^2}.
\end{equation}
Here $\braket{H^2}=\braket{\phi(\vec{\theta}(t))|H^2|\phi(\vec{\theta}(t))}$. Therefore, by measuring $A^R$, $C^I$, and $\braket{H^2}$, we can calculate the distance to verify the variational algorithm. \yx{This works similarly for the modified equation that takes account of the global phase.}

\subsubsection{The time-dependent variational principle}
The Schr\"odinger equation can be obtained from the Lagrangian  $L=\braket{\psi(t)|(d/d t+iH)|\psi(t)}$ \yx{or more precisely $iL$ so that the Lagrangian is real. As the imaginary term $i$ does not affect the evolution equation, we omit it for simplicity.}  Replacing $\ket{\psi(t)}$ with $\ket{\phit}$, the  Lagrangian becomes
\begin{equation}
\begin{aligned}
	L &= \left\langle{{\phit}\bigg|\left(\frac{d}{d t}+iH\right)}\bigg|{\phit}\right\rangle.
\end{aligned}
	\end{equation}
For complex $\theta_i$, parameters are $\theta_i$, ${\theta}_i^*$, $\dot{\theta}_i$, and $\dot{\theta}_i^*$. After applying the Euler-Lagrange equation, the evolution of parameters can be derived and shown to coincide with Eq.~\eqref{Eq:pureUnitaryComplex}. 
For real $\theta_i$, parameters of the Lagrangian are $\theta_i$ and $\dot{\theta}_i$, and the evolution of parameters are
\begin{equation}\label{Eq:pureImagSolution}
		\sum_j A_{i,j}^I \dot\theta_j = -C_i^R,
\end{equation}
which can be obtained by taking the imaginary part of Eq.~\eqref{Eq:pureUnitaryComplex}. Here $A^I_{i,j}$ is the imaginary part of $A_{i,j}$ and $C_i^R$ is the real part of $C_i$. As shown in Sec.~\ref{Sec:equivalence}, this evolution equation is equivalent to Eq.~\eqref{Eq:pureUnitaryComplex} when the trial state $\ket{\phit}$ is powerful enough to represent the target state $\ket{\psi(t)}$. However, as we take the imaginary part of the $A$ matrix, the $A^I$ matrix is more likely to be singular, making the evolution unstable. 
\yx{Note that there also exists other ways of choosing the Lagrangian, such as the one considered in Ref.~\cite{PhysRevB.88.075133},
$$L = \frac{i}{2}\left(\bra{\psi} \frac{d\ket{\psi}}{d t} - \frac{d\bra{\psi}}{d t}\ket{\psi}\right)  - \braket{\psi| H |\psi}.$$ 
Nevertheless, it is not hard to verify that it leads to the same evolution equation as Eq.~\eqref{Eq:pureImagSolution}. }

\subsection{Implementation}
In variational classical simulation, the $A$ matrix and the $C$ vector are calculated classically. Here we show how to evaluate $A$ and $C$ with a quantum circuit introduced by \cite{Li2017}. Suppose the parameterised trial state is prepared by $\ket{\phi(\vec{\theta})}  = R\ket{\bar{0}}$ with $R = R_{N}(\theta_N)\dots R_{k}(\theta_k)\dots R_{1}(\theta_1)$. In conventional quantum computing, each unitary gate involves a small subset of qubits, typically one or two. Therefore, the derivative of the unitary can be efficiently decomposed via
\begin{eqnarray}
\frac{\partial R_{k}}{\partial \theta_{k}} = \sum_{i} f_{k,i} R_{k} \sigma_{k,i},
\label{eq:expansion}
\end{eqnarray}
where $\sigma_{k,i}$ are also unitary operators and $f_{k,i}$ are complex coefficients. For most single- and two-qubit-gates $R_{k}$, we find that there are only one or two terms in this expression, and $\sigma_{k,i}$ is also a one-qubit or two-qubit gate. For example, when $R_k(\theta_k)=e^{-i\theta_k\sigma/2}$ with a one- or two-qubit Hermitian operator $\sigma$, we have ${\partial R_{k}(\theta_k)}/{\partial \theta_k} = -i/2\cdot \sigma         \cdot e^{-i\theta_k\sigma/2}$ with $f_k=-i/2$ and $\sigma_k = \sigma         $ in Eq.~\eqref{eq:expansion}.

Using the expression (\ref{eq:expansion}), we rewrite the derivative of the state as
\begin{eqnarray}
\frac{\partial \ket{\phit}}{\partial \theta_{k}} = \sum_{i} f_{k,i} R_{k,i} \ket{\bar{0}},
\end{eqnarray}
where
\begin{eqnarray}
R_{k,i}= R_{N} R_{N-1} \cdots R_{k+1} R_{k} \sigma_{k,i} \cdots R_{2} R_{1}.
\end{eqnarray}
Then, based on the definition of the Hamiltonian, the elements of $A$ and $C$ can be expressed as
\begin{eqnarray}
A_{k,q} = \sum_{i,j} \left(
f^*_{k,i}f_{q,j} \bra{\bar{0}} R^\dag_{k,i} R_{q,j} \ket{\bar{0}}
\right),
\label{eq:M}
\end{eqnarray}
and
\begin{eqnarray}
C_{k} = \sum_{i,j} \left(
f^*_{k,i}h_{j} \bra{\bar{0}} R^\dag_{k,i}\sigma_{j}R \ket{\bar{0}}
 \right).
\label{eq:V}
\end{eqnarray}

\begin{figure*}[t]
\centering
\includegraphics[width=0.75\linewidth]{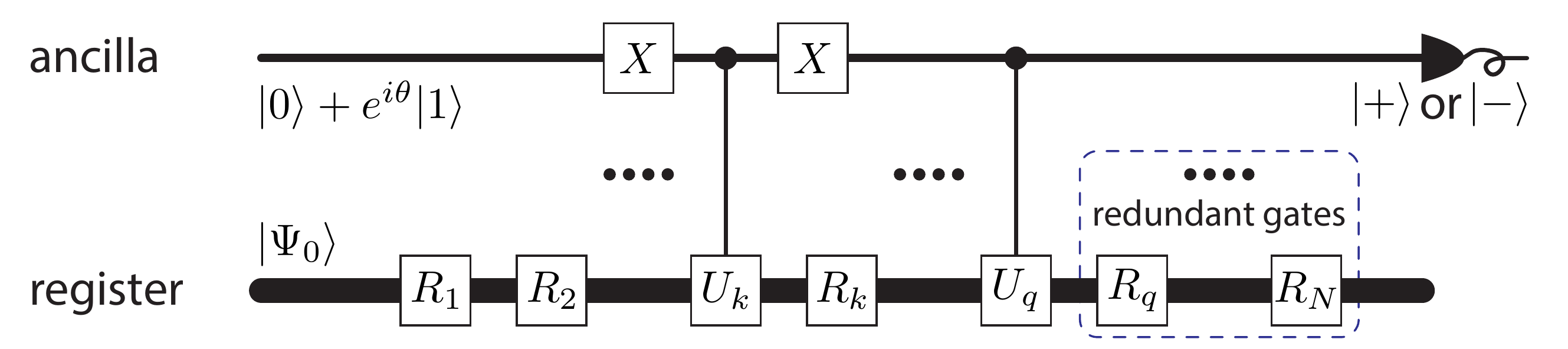}
\caption{
Quantum circuit for the evaluation of coefficients in the variational pure-state simulator (from Ref.~\cite{Li2017}). 
The ancillary qubit is initialised in the state $(\ket{0}+e^{i\theta}\ket{1})/\sqrt{2}$ and measured in the $\ket{\pm} = (\ket{0} \pm \ket{1})/\sqrt{2}$ basis.
The probability $P_{+}$ that the qubit is in the state $\ket{+}$ is $(\Re \left( e^{i\theta} \bra{\bar{0}} U \ket{\bar{0}} \right)+1)/2$, where $U = R_{1}^\dag \cdots U_{k}^\dag R_{k}^\dag \cdots R_{N}^\dag R_{N} \cdots R_{q} U_{q} \cdots R_{1}$,  Here, $U_{k}$ is one of $\sigma_{k,i}$, and $U_{q}$ is one of $\sigma_{q,j}$ or $\sigma_{j}$ (By taking $q = N+1$, $\sigma_{j}$ is put on the left side of $R_{N}$ in the product), assuming that $k < q$. We would like to remark that this circuit is a variant of the circuit proposed in Ref.~\cite{ekertcircuit02}. This circuit contains $N$ gates on the register, two flip gates ($X$) on the ancillary qubit, and two controlled unitary gates between the ancillary qubit and the register. Note that gates on the register after the second controlled unitary gate can be omitted. Usually, $U_{k}$ and $U_{q}$ are unitary operators on only one or two register qubits rather than the entire register. This circuit can be further reduced to a direct measurement on the register qubits without the ancilla \cite{mitarai2018methodology}. 
}
\label{fig:circuit}
\end{figure*}

The real and imaginary part of expressions (\ref{eq:M}) and (\ref{eq:V}) are in the form
$$
a \Re \left( e^{i\theta} \bra{\bar{0}} U \ket{\bar{0}} \right),
$$
where the amplitude $a$ and phase $\theta$ are determined by the coefficients, and $U$ is a unitary operator either $R^\dag_{k,i}R_{q,j}$ or $R^\dag_{k,i}\sigma_{j}R$. Such a term can be evaluated with the quantum circuit shown in Fig.~\ref{fig:circuit}. This circuit needs an ancillary qubit initialised in the state $(\ket{0}+e^{i\theta}\ket{1})/\sqrt{2}$ and a register initialised in the state $\ket{\Psi_0}$. The ancillary qubit is measured after a sequence of unitary operations on the register and two controlled unitary operations, where the ancillary qubit is the control qubit. The value is given by $\Re \left( e^{i\theta} \bra{\bar{0}} U \ket{\bar{0}} \right) = 2P_{+} - 1$, where $P_{+}$ is the probability that the qubit is in the state $\ket{+} = (\ket{0}+\ket{1})/\sqrt{2}$. Recently, \cite{mitarai2018methodology} also showed a way to measure $A$ and $C$ without the ancillary qubit, making the realisation more friendly to near-term quantum hardware.

\section{Equivalence of the three variational principles}\label{Sec:equivalence}
With complex parameters of the trial state, 
the three variational principles lead to the same evolution equation of parameters. We will see that this is also true for general stochastic evolution and imaginary time evolution of mixed states.
However, when considering real parameters as in quantum simulation, we can get three different equations \eqref{Eq:pureUnitaryComplex}, \eqref{Eq:pureRealSolution}, and \eqref{Eq:pureImagSolution} for the evolution of parameters. 
With classical ansatz, a necessary condition for the equivalence of the three equations is discussed by \cite{broeckhove1988equivalence}. That is these three equations are equivalent when for any parameter $\theta_i$, there always exists a parameter $\theta_j$, such that 
\begin{equation}
\frac{\partial \ket{\phit}}{\partial \theta_i} = i\frac{\partial \ket{\phit}}{\partial \theta_j}.
\end{equation}

However, such a condition is hard to be satisfied in quantum simulation with real parameters. Generally, when $\ket{\phit}$ can uniquely represent any state $\ket{\psi(t)}$ at time $t$, Eq.~\eqref{Eq:pureUnitaryComplex} and Eq.~\eqref{Eq:pureRealSolution} both give real solutions and are  also equivalent.
The necessary and sufficient condition for the equivalence of Eq.~\eqref{Eq:pureUnitaryComplex} and Eq.~\eqref{Eq:pureRealSolution} only requires the solution of Eq.~\eqref{Eq:pureUnitaryComplex} to be real.

Broadly speaking we may expect that Eq.~\eqref{Eq:pureRealSolution} and Eq.~\eqref{Eq:pureImagSolution} each suffice to define the evolution of our parameters. Practically, there are differences however.  
In particular, we can consider a single qubit system with Hamiltonian $\sigma_x$ and initial state $\ket{0}$.  Suppose the trial state is chosen to be $e^{-i\theta \sigma_x}\ket{{0}}$, which can exactly represent the evolution of the state. For such a simple case, the evolution equation \eqref{Eq:pureImagSolution} based on the time-dependent variational principle cannot work as $A^I$ and $C^R$ are always zero; nevertheless, Eq.~\eqref{Eq:pureUnitaryComplex} and Eq.~\eqref{Eq:pureRealSolution} can both simulate the evolution. 
In fact, as the diagonal elements of $A$ are in general real constants, the diagonal elements of $A^R$ and $A^I$ are real constants and zero, respectively. As the diagonal elements of  $A^I$ are all zero, it makes the inverse of $A^I$ unstable when the off-diagonal elements are also close to 0. This suggests that the evolution equation \eqref{Eq:pureImagSolution} may not be the ideal choice for VQS.

In the following, we will re-derive the evolution equations from the mixed state perspective. Surprisingly,  we will show that Eq.~\eqref{Eq:pureImagSolution} from the time-dependent variational principle cannot be obtained in the mixed state case, whereas a variant of Eq.~\eqref{Eq:pureRealSolution} can be consistently derived. Therefore, our work suggests that  McLachlan's principle is the most consistent variational principle in variational quantum simulation. \yx{We also summarise the evolution equations of parameters under different conditions in Table~\ref{tab:results}. }

\begin{table*}[t]\footnotesize
\centering
  \caption{\yx{Summary of evolution equation of parameters under different conditions. We consider real/imaginary time dynamics for pure/mixed states with complex/real parameters and three different variational principles. TDVP: time-dependent variational principle. The evolution equation is the same for complex parameters of the three variational principles. 
  As pure state unitary dynamics is a special case of the mixed state open system dynamics, the evolution equation for mixed states also works for pure states, which further corrects the global phase problem for pure states as discussed in  Sec.~\ref{Sec:globalphase}. The definition of the matrices are: 
  $A_{i,j} = \frac{\partial \bra{\phi(\vec{\theta}(t))}}{\partial \theta_i}\frac{\partial \ket{\phi(\vec{\theta}(t))}}{\partial \theta_j}$ and $C_i= \frac{\partial \bra{\phi(\vec{\theta}(t))}}{\partial \theta_i}H\ket{\phi(\vec{\theta}(t))}$ are defined in Eq.~\eqref{Eq:defAandC};  $
		M_{i,j} = \tr\left[\left(\frac{\partial \rhot}{\partial \theta_i}\right)^\dag\frac{\partial \rhot}{\partial \theta_j}\right]$, $V_i = \tr\left[\left(\frac{\partial \rhot}{\partial \theta_i}\right)^\dag\ml\right]$ are defined in Eq.~\eqref{eq:MVmatrixFirst}. $
		C_i' = \frac{\partial \bra{\phi(\vec{\theta}(\tau))}}{\partial \theta_i}(H-E_\tau)\ket{\phi(\vec{\theta}(\tau))}$ is defined in Eq.~\eqref{Eq:defCprime}; $
	Y_i=-\mathrm{Tr}\left[ \frac{\partial \rhot}{\partial \theta_i}  \{H, \rhot\} \right]$ is defined in Eq.~\eqref{Eq:Yireduced}.}
		}
\begin{tabular}{ccccc}
\specialrule{.1em}{.05em}{.05em} 
Dynamics&Pure/mixed states&Parameters&Variational principles&Evolution equations\\
\specialrule{.1em}{.2em}{.2em} 
\multirow{8}{*}{Real}&\multirow{4}{*}{Pure}&Complex&-&$\sum_j A_{i,j}\dot{\theta}_j = -iC_i$ \eqref{Eq:pureUnitaryComplex}\\
\cdashline{3-5}
&&\multirow{3}{*}{Real}&Dirac and Frenkel&$\sum_j A_{i,j}\dot{\theta}_j = -iC_i$ \eqref{Eq:pureUnitaryComplex}\\
&&&MacLachlan&$\sum_j A^R_{i,j}\dot{\theta}_j = C_i^I$ \eqref{Eq:pureRealSolution}\\
&$\frac{d \ket{\psi(t)}}{d t} = -iH\ket{\psi(t)}$ \eqref{Schrodinger}&&TDVP&$\sum_j A^I_{i,j}\dot{\theta}_j = -C_i^R$ \eqref{Eq:pureImagSolution}\\
\cline{2-5}
&\multirow{4}{*}{Mixed}&Complex&-&$\sum_j M_{i,j}\dot{\theta}_j=V_i$ \eqref{Eq:MVequation}\\
\cdashline{3-5}
&&\multirow{3}{*}{Real}&Dirac and Frenkel&$\sum_j M_{i,j}\dot{\theta}_j=V_i$ \eqref{Eq:MVequation}\\
&&&MacLachlan&$\sum_j M_{i,j}\dot{\theta}_j=V_i$ \eqref{Eq:MVequation}\\
&$\frac{d \rho}{d t} = \ml$ \eqref{Eq:stochasticdef}  &&TDVP&-\\
\cline{1-5}
\multirow{8}{*}{Imag}&\multirow{4}{*}{Pure}&Complex&-&$\sum_j A_{i,j}\dot{\theta}_j = -C_i'$ \eqref{Eq:imagpureevolution}\\
\cdashline{3-5}
&&\multirow{3}{*}{Real}&Dirac and Frenkel&$\sum_j A_{i,j}\dot{\theta}_j = -C_i'$ \eqref{Eq:imagpureevolution}\\
&&&MacLachlan&$\sum_j A_{i,j}^R\dot{\theta}_j = -C_i^R$ \eqref{Eq:imaginaryevolAC}\\
&$\frac{d \ket{\psi(\tau)}}{d \tau} = -(H-E_\tau)\ket{\psi(\tau)}$ \eqref{eq:Schrodinger2}&&TDVP&-\\
\cline{2-5}
&\multirow{4}{*}{Mixed}&Complex&-&$\sum_j M_{i,j} \dot{\theta}_j = Y_i$ \eqref{Eq:mixedImaginary}\\
\cdashline{3-5}
&&\multirow{3}{*}{Real}&Dirac and Frenkel&$\sum_j M_{i,j} \dot{\theta}_j = Y_i$ \eqref{Eq:mixedImaginary}\\
&&&MacLachlan&$\sum_j M_{i,j} \dot{\theta}_j = Y_i$ \eqref{Eq:mixedImaginary}\\
&$\frac{ d \rho (\tau) }{d \tau} = -\left( \{H, \rho(\tau)\} - 2\braket{H} \rho(\tau) \right)$ \eqref{eq:imagerho}&&TDVP&-\\
\specialrule{.1em}{.2em}{.2em} 
\end{tabular}
\label{tab:results}
\end{table*}

\section{{Real time evolution: open quantum systems}}\label{Sec:real}
In this section, we extend VQS to mixed states under general stochastic evolutions, which can describe open quantum systems and noisy quantum hardware. 
Note that this approach differs from that in Ref.~\cite{suguru18theory} where we can variationally simulate the stochastic mater equation.


\subsection{General evolution}
The general stochastic evolution of an open quantum system can be modelled by
\begin{equation}\label{Eq:stochasticdef}
	\frac{d \rho}{d t} = \ml,
\end{equation}
where the superoperator $\ml$ can be decomposed as
\begin{equation}
	\ml = \sum_i g_iS_i\rho T_i^\dag,
\end{equation}
with unitary operators $S_i$ and $T_i$, and coefficients $g_i$. Suppose $\rho = \rhot$ is a parameterised state, we show how to transform the evolution of $\rho$ into the evolution of parameters $\vec{\theta}$. 
Here, we mainly focus on the evolution of parameters and refer to Sec.~\ref{Sec:implementation} for the design of the mixed states ansatz with unitary circuits.

\subsubsection{The Dirac and Frenkel variational principle}
Let $\rho = \rhot$, the general stochastic evolution becomes
\begin{equation}
	\sum_i \frac{\partial \rhot}{\partial \theta_i}\dot{\theta}_i = \ml. 
\end{equation}
Because $\frac{d \rhot}{d t} = \sum_i \frac{\partial \rhot}{\partial \theta_i}\dot{\theta}_i$ is in the tangent subspace of $\{\frac{\partial \rhot}{\partial \theta_i}\}$, we can project the  equation onto this tangent subspace by following the Dirac and Frenkel variational principle, 
\begin{equation}
	\tr\left[(\delta\rhot)^\dag\left(\frac{d \rho}{d t}-\ml\right)\right] = 0.
\end{equation}
The evolution of parameters is
\begin{equation}\label{Eq:MVequation}
\sum_j M_{i,j}\dot{\theta}_j=V_i,
\end{equation}
with elements of $M$ and $V$ defined by
\begin{equation}\label{eq:MVmatrixFirst}
	\begin{aligned}
		M_{i,j} &= \tr\left[\left(\frac{\partial \rhot}{\partial \theta_i}\right)^\dag\frac{\partial \rhot}{\partial \theta_j}\right],\\
		V_i &= \tr\left[\left(\frac{\partial \rhot}{\partial \theta_i}\right)^\dag\ml\right].
	\end{aligned}
\end{equation}
In classical simulation, $\vec{\theta}$ can be complex and $ \left(\frac{\partial \rhot}{\partial \theta_i}\right)^\dag\neq\frac{\partial \rhot}{\partial \theta_j}$. In this case, $M$, $V$, and the solution $\dot\theta$ can all be complex. However, considering quantum simulation with real parameters $\theta$, we have $\left(\frac{\partial \rhot}{\partial \theta_i}\right)^\dag=\frac{\partial \rhot}{\partial \theta_j}$. Therefore, $M$ and $V$ are real and the solution $\dot{\theta}$ is also real. This is different from the pure state case where even if parameters are real, the solution $\dot{\theta}$ can still be complex. Each term of $M$ and $V$ can be calculated with a quantum circuit that is discussed in Sec.~\ref{Sec:implementation}.

\subsubsection{McLachlan's variational principle}
We can also minimise the distance between the evolution of the trial state and the true evolution, 
\begin{equation}
 	\delta \|{d \rho}/{d t} -\ml\|=0,
\end{equation}
\yxx{where $\|\rho\| = \sqrt{\tr[\rho^2]}$ is the $l_2$ norm of matrices.} 
With both complex and real parameters,  we can get the same result with the Dirac and Frenkel variational principle. Meanwhile, after solving the derivatives of $\dot\theta$, we can also get the distance between the true evolution and the evolution of the trial state by calculating
\begin{equation}
	\|{d \rho}/{d t} -\ml\|^2 = \sum_{i,j}M_{i,j}\dot\theta_i\dot\theta_j - 2\sum_i V_i\dot\theta_i + \tr[\ml^2].
\end{equation}
Therefore, by measuring $M$, $V$, and $\tr[\ml^2]$, we can calculate the distance to verify the variational algorithm. 

\subsubsection{The time-dependent variational principle}
\yx{In general, Hamiltonians for open systems may not be well defined due to the dissipation. Here, we construct the Lagrangian as follows, 
\begin{equation}
	L = \tr[\rhot^\dag({d \rhot}/{d t}  -\ml)],
\end{equation}
so that after the Euler-Lagrange equation, we can recover the evolution equation defined in Eq.~\eqref{Eq:stochasticdef}. Here, we assume that $\theta$ are complex parameters so that $\rho^\dag$ and $\rho$ can be also regarded to be independent. 
Applying the Euler-Lagrange equation to $\theta^*$ gives the same evolution Eq.~\eqref{Eq:MVequation}.
However, when parameters are real with $\theta=\theta^*$, we also have $\rho^\dag=\rho$ and the Euler-Lagrange equation cannot lead to Eq.~\eqref{Eq:stochasticdef} or any evolution equation of the parameters. We refer to  Appendix~\ref{Sec:appendix2} for details.}\\

From the above result, we find that the three variational principles are equivalent for complex parameters. However, when parameters are real, the Dirac and Frenkel and the McLachlan variational principles are also equivalent, while the time-dependent variational principle cannot lead to a nontrivial evolution of parameters. Therefore, the McLachlan variational principle always gives a consistent real evolution of parameters in VQS. 

\subsection{Reduction to unitary evolution}\label{Sec:globalphase}
\yx{Note that the unitary evolution of Schr\"odinger's equation is a special case of the stochastic evolution defined in Eq.~\eqref{Eq:stochasticdef}. Therefore, it would be interesting to verify whether the evolution equations of parameters for Schr\"odinger's equation can be consistently obtained from the more general scenario. }
First, we consider unitary evolution of mixed input states. Suppose the initial state is $\rho(0)$, the evolution under Hamiltonian $H$ is governed by the von Neumann equation with $\mathcal{L}(\rho) = -i[H, \rho(t)]$,
\begin{equation}
	\frac{d \rho(t)}{d t} = -i[H, \rho(t)]. 
\end{equation}
In this case, $M$ is defined in Eq.~\eqref{eq:MVmatrixFirst} and $V$ is explicitly given by
\begin{equation}
	\begin{aligned}
		V_i &= -i\tr\left[\left(\frac{\partial \rhot}{\partial \theta_i}\right)^\dag[H, \rhot]\right].
	\end{aligned}
\end{equation}

\begin{figure}[t]\centering
  \includegraphics[width=10cm]{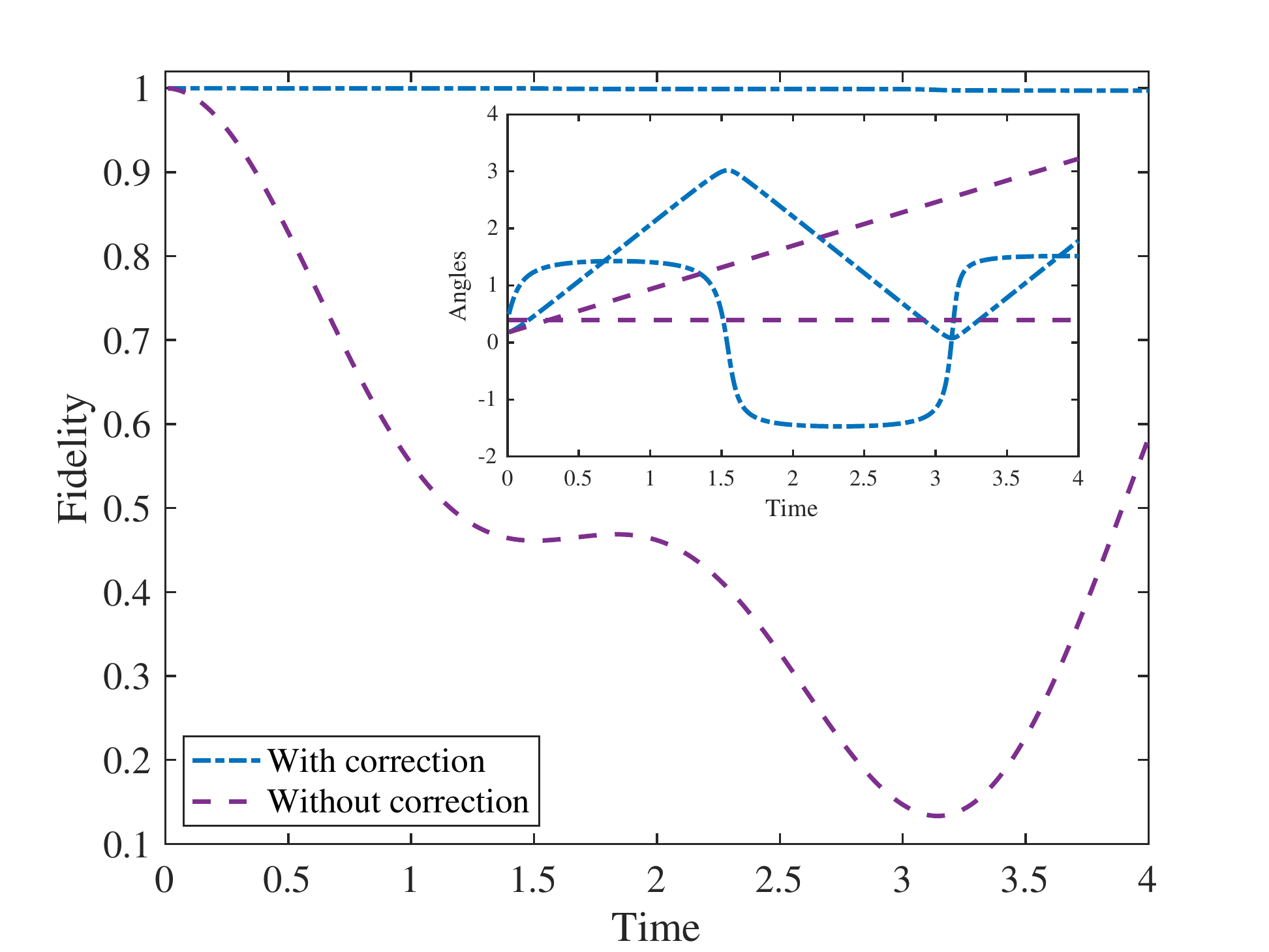}
  \caption{Comparison between the evolution equations Eq.~\eqref{Eq:mixedunitary} and Eq.~\eqref{Eq:pureRealSolution} for real time evolution with Hamiltonian $\sigma_y$ and trial state $\ket{\phit} = R_z(\theta_z)R_x(\theta_x)\ket{0}$. Here, we consider randomly initialised state ($\theta_1=0.1734$ and $\theta_2 = 0.3909$ for the shown results) and find that the evolution under Eq.~\eqref{Eq:pureRealSolution} generally fails to simulate the true evolution even if the ansatz can represent all pure qubit state. 
The Y-axis is the fidelity of the state under variational simulation to the state  under the exact evolution. Dash-dot line corresponds to Eq.~\eqref{Eq:mixedunitary}; Dashed line corresponds to Eq.~\eqref{Eq:pureRealSolution}. The evolution of parameters are shown in the inset.
  }\label{Fig:compare}
\end{figure}

Suppose $\rhot=\ket{\phit}\bra{\phit}$ is a pure state, it becomes the Schr\"odinger equation and we have
\begin{equation}\label{Eq:mixedunitary}
	\begin{aligned}
		M_{i,j} &= A_{i,j}^R+ \frac{\partial \bra{\phi(\vec{\theta}(t))}}{\partial \theta_i}\ket{\phit}\frac{\partial \bra{\phi(\vec{\theta}(t))}}{\partial \theta_j}\ket {\phit},\\ 
		V_i &= C_i^I+i\frac{\partial \bra{\phi(\vec{\theta}(t))}}{\partial \theta_i}\ket{\phit}\bra{\phit}H\ket{\phi(\vec{\theta}(t))}.
	\end{aligned} 
\end{equation}
The evolution equation obtained from directly applying the McLachlan's principle to pure states with unitary evolution is $\sum_j A^R_{i,j}\dot{\theta}_j = C_i^I$, as given in Eq.~\eqref{Eq:pureRealSolution}. 
Compared to the $A^R$ matrix and $C^I$ vector, there are additional terms that estimate the overlap between the trial state and the derivative of the trial state in $M$ and $V$. 
The terms $\frac{\partial \bra{\phi(\vec{\theta}(t))}}{\partial \theta_i}\ket{\phit}$ can be measured with the circuit in Fig.~\ref{fig:circuit} and the average energy $\bra{\phit}H\ket{\phi(\vec{\theta}(t))}$ can be directly measured.
As we have discussed in Sec.~\ref{Sec:preliminary}, the additional terms take account of the global phase difference between the trial state and the target evolution state. Consider a trial state $e^{i\theta_{0}}\ket{\phit}$ with $N$ parameters in $\ket{\phit}$ and an extra parameter $\theta_{0}$ for the global phase. Using Eq.~\eqref{Eq:pureRealSolution}, the evolution of the first $N$ parameters of $e^{i\theta_{0}}\ket{\phit}$ is exactly given by Eq.~\eqref{Eq:mixedunitary}. Because the global phase $\theta_{0}$ is irrelevant to any measurement, we can thus only evolve the first $N$ parameters with Eq.~\eqref{Eq:mixedunitary}. However, the extra parameter $\theta_{0}$ is important for the global phase alignment between the trial state and the exact state as we confirm with the following example. 

Here, we consider a single-qubit state evolution to illustrate the difference between  Eq.~\eqref{Eq:mixedunitary} and Eq.~\eqref{Eq:pureRealSolution}. Suppose the trial state is prepared by $\ket{\phit} = R_z(\theta_z)R_x(\theta_x)\ket{0}$ with two parameters  $\theta_z$ and $\theta_x$. Here, $R_x(\theta) = e^{-i\theta\sigma_x/2}$ and $R_z(\theta) = e^{-i\theta\sigma_z/2}$, with Pauli matrices $\sigma_x$ and $\sigma_z$. Suppose the system Hamiltonian is the Pauli matrix $\sigma_y$, we compare the simulation of the time evolution $e^{-i\sigma_y t}$ with randomly initialised state, as shown in Fig.~\ref{Fig:compare}. We find that the evolution under Eq.~\eqref{Eq:pureRealSolution} generally fails to simulate the exact real time dynamics. Nevertheless, as the trial state can represent all pure qubit state up to a global phase, the evolution under Eq.~\eqref{Eq:mixedunitary} can indeed simulate the true time evolution $e^{-i\sigma_y t}$. Therefore, Eq.~\eqref{Eq:mixedunitary} is recommended for variational simulation of real time unitary dynamics of pure states.

\section{Imaginary time evolution}\label{Sec:imaginary}
In this section, we show that the variational principles can be also applied to the simulation of imaginary time evolution. 
\subsection{Pure states}
We first focus on the pure state case studied in Ref.~\cite{mcardle2018variational}.
Replacing real time with imaginary time $\tau = it$, the state at time $\tau$ is $\ket{\psi(\tau)} = \frac{e^{-H\tau}\ket{\psi(0)}}{\sqrt{\bra{\psi(0)}e^{-2H\tau}\ket{\psi(0)}}}$, and the Wick-rotated Schr\"odinger equation is,
\begin{equation}\label{eq:Schrodinger2}
	\frac{d \ket{\psi(\tau)}}{d \tau} = -(H-E_\tau)\ket{\psi(\tau)},
\end{equation}
where $E_\tau = \braket{{\psi(\tau)}|H|{\psi(\tau)}}$ is the expected energy at time $\tau$.
Consider a normalised parametrised trial state $\ket{\phitau}$, the imaginary time evolution of the Schr\"odinger equation on the trial state space is
\begin{equation}\label{diffEq2}
	\sum_i \frac{\partial \ket{\phitau}}{\partial \theta_i}\dot{\theta_i}=-(H-E_\tau)\ket{\phitau}.
\end{equation}
Therefore, we can evolve parameters to simulate the unphysical imaginary time evolution.\\

\subsubsection{The Dirac and Frenkel variational principle}
Applying the Dirac and Frenkel variational principle, we have 
\begin{equation}
	\left\langle{ \delta\phitau\bigg|\left(\frac{d}{d \tau}+H-E_\tau\right)\bigg|\phi(\vec{\theta}(\tau))}\right\rangle = 0,
\end{equation}
and the evolution of parameters is simplified to
\begin{equation}\label{Eq:imagpureevolution}
	\sum_j A_{i,j}\dot{\theta}_j = -C_i',
\end{equation}
with
\begin{equation}\label{Eq:defCprime}
	\begin{aligned}
		C_i' &= \frac{\partial \bra{\phi(\vec{\theta}(\tau))}}{\partial \theta_i}(H-E_\tau)\ket{\phi(\vec{\theta}(\tau))}.
	\end{aligned}
\end{equation}
Similar to the real time evolution, even if the parameter $\theta_j$ is real, the solution of its derivative $\dot \theta_j$ may not be real as $A$ and $C'$ are complex.

\subsubsection{McLachlan's variational principle}
Applying the McLachlan's variational principle \cite{McLachlan} to imaginary time evolution, we have
 \begin{equation}
 	\delta \|({d}/{d \tau} + H-E_\tau)\ket{\phitau)}\|=0.
 \end{equation}
When $\dot{\theta}_i$ is complex, the evolution of parameters is the same as Eq.~\eqref{Eq:imagpureevolution}.
When $\dot{\theta}_i$ can only take real values, the evolution becomes
\begin{equation}\label{Eq:imaginaryevolAC}
	\sum_j A^R_{i,j}\dot{\theta}_j = -C_i^R.
\end{equation}
Due to the normalisation requirement for the trial state $\ket{\phitau}$, i.e., $\bra{\phi(\vec{\theta}(\tau))}\phitau\rangle = 1$, we have $\Re\left(\frac{\partial \bra{\phit}}{\partial \theta_i}E_\tau\ket{\phit}\right) = 0$ and therefore $C_i^R=C_i'^R$.

\subsubsection{Time-dependent variational principles}
The Lagrangian for the Schr\"odinger equation is
\begin{equation}
\begin{aligned}
	L &=\left\langle{\psi}\bigg|\frac{d}{d \tau}+H\bigg|\psi\right\rangle+\braket{\psi|H|\psi}(1-\braket{\psi|\psi}),
\end{aligned}
\end{equation}
\yx{which reproduces the imaginary time evolution equation defined inn Eq.~\eqref{eq:Schrodinger2}.}
When $\ket{\psi} = \phit$ has complex $\theta$, we get the evolution of Eq.~\eqref{Eq:imagpureevolution}.
Suppose $\theta$ is real, the evolution becomes
\begin{equation}
		i\sum_j A_{i,j}^I \dot\theta_j = C_i^R,
\end{equation}
where, $A^I_{i,j}$ and $C_i^R$ are the imaginary and real parts of $A_{i,j}$ and $C_i$, respectively. The solution is imaginary and hence not physical. \\

To summarise, when parameters are complex, the three variational principles are still equivalent. However, considering real parameters, only the McLachlan's variational principle can guarantee a real solution of the evolution equation.

\subsection{Mixed state}
Now, we consider the case that the input state is a mixed state. 
Under the imaginary time evolution of Hamiltonian $H$, the state $\rho(\tau)$ at imaginary time $\tau$ is 
\begin{equation}\begin{aligned}
\rho(\tau)= \frac{e^{-H \tau} \rho(0) e^{-H \tau}}{\mathrm{Tr}[e^{-2 H \tau} \rho(0)]},  
\end{aligned} \end{equation}
where $\rho(0)$ is the initial state. Equivalently, $\rho(\tau)$ follows the time derivative equation 
\begin{equation}\label{eq:imagerho}
	\frac{ d \rho (\tau)}{d \tau}  = -\bigg( \{H, \rho(\tau)\} - 2\braket{H} \rho(\tau) \bigg), 
\end{equation}
where $\{H, \rho(\tau)\} = H\rho(\tau)+\rho(\tau)H$.
Consider a parametrised state $\rhot$, the evolution becomes 
\begin{equation}\begin{aligned}
\sum_i \frac{\partial \rho (\vec{\theta})}{\partial \theta_i} \dot{\theta}_i = -\bigg( \{H, \rhot\} - 2\braket{H} \rhot \bigg). 
\end{aligned} \end{equation}

\subsubsection{The Dirac and Frenkel variational principle}
To get the evolution of parameters, we project the equation onto $\delta \rho (\vec{\theta}) $ according to the  Dirac and Frenkel variational principle. Therefore,  we have 
 \begin{equation}\begin{aligned}
&\mathrm{Tr}\bigg[\delta \rho (\vec{\theta}) \bigg( \sum_i \frac{\partial \rho (\vec{\theta})}{\partial \theta_i} \dot{\theta}_i  +\{H, \rhot\} - 2\braket{H} \rhot \bigg) \bigg]
\end{aligned} \end{equation}
equal zero and the solution is 
\begin{equation}\begin{aligned}\label{Eq:mixedImaginary}
\sum_j M_{i,j} \dot{\theta}_j = Y_i,
\end{aligned} \end{equation}
with $M$ defined in Eq.~\eqref{eq:MVmatrixFirst} and $Y$ given by 
\begin{equation}\begin{aligned}
Y_i &= -\mathrm{Tr}\bigg[ \bigg( \frac{\partial \rho (\vec{\theta})}{\partial \theta_i} \bigg)^\dag\bigg( \{H, \rhot\} -2 \braket{H} \rhot \bigg) \bigg].
\end{aligned} \end{equation}
Note that when parameters $\vec{\theta}$ are real, $M$ and $Y$ are also real, and we have
\begin{equation}\label{Eq:Yireduced}
	Y_i=-\mathrm{Tr}\bigg[ \frac{\partial \rhot}{\partial \theta_i}  \{H, \rhot\} \bigg].
\end{equation}
In this case, the solution $\vec{\dot{\theta}}$ is also real note that this is opposite to the pure state case.

\subsubsection{MacLachlan's variational principle}
We can also minimise the distance between the evolution via parameters and the evolution,
\begin{equation}\begin{aligned}
\delta \| d \rho/ d \tau +(\{H, \rho\} - 2 \braket{H} \rho) \| =0. 
\end{aligned} \end{equation}
The evolution to $\dot{\vec{\theta}}$ is the same as Eq.~\eqref{Eq:mixedImaginary}. Similarly, when $\dot{\vec{\theta}}$ is real, $Y_i$ is reduced to Eq.~\eqref{Eq:Yireduced} and we always have real solutions. 

\subsubsection{The time-dependent variational principle}
\yx{
The Lagrangian for the imaginary time evolution is 
\begin{equation}
	\begin{aligned}
L=\mathrm{Tr} \bigg[\rhot^\dag \bigg(\frac{d \rhot}{d t} +\{H, \rhot\} -2\braket{H} \rhot\bigg) \bigg].
	\end{aligned}
\end{equation}
When $\vec{\theta}$ is complex, we can regard $\rhot^\dag$ and $\rhot$ as independent parameters and hence recover the equation defined in Eq.~\eqref{eq:imagerho}. Applying the Euler-Lagrangian equation to $\theta^*$, we can obtain the evolution the same as in Eq.~\eqref{Eq:mixedImaginary}.
Instead, when $\vec{\theta}$ is real, we cannot regard $\rhot^\dag$ and $\rhot$ as independent parameters and thus we cannot recover Eq.~\eqref{eq:imagerho} or obtain the evolution equation of the parameters ${\vec{\theta}}$. }\\

As a special case, suppose $\rhot$ is a pure state $\ket{\phit}$, we have the same $M_{i,j}$ as in real time evolution given in Eq.~\eqref{Eq:mixedunitary} and $Y_i = -C_i^R$.

\subsection{Variational Gibbs state preparation}
Evolving the maximally mixed state $I_d/d$ with imaginary time $\tau$ can be used to prepare the thermal state $\rho(T) = e^{-H/T}/\tr[e^{-H/T}]$ with temperature $T = 1/2\tau$. Here, $d$ is the dimension of the system. \yx{Variational classical simulation of finite-temperature Gibbs state with matrix product states has been introduced in Ref.~\cite{Verstraete04}.} For variational quantum simulation, we cannot simply apply a variational unitary circuit to the maximally mixed state and change parameters in the unitary circuit realise the simulation. This is because $I_d/d$ is invariant under unitary, $U(\theta)\cdot I_d/d \cdot U(\theta)^\dag = I_d/d$. Instead, we can input a maximally entangled state $\ket{\Phi}_d = 1/\sqrt{d}\sum_i\ket{ii}_{AE}$ of system $AE$ and evolve the whole system with Hamiltonian $H\otimes I_d$ under imaginary time $\tau$. Then, the state of system $A$ at time $\tau$ will be the thermal state with temperature $T = 1/2\tau$.

\section{Implementation}\label{Sec:implementation}
In this section, we first show how to realise pure and mixed trial states with quantum circuits possibly assisted with post-selection on measurement outcomes.  Then we show how to measure each term of $M$, $V$, and $Y$ for simulating general evolutions of mixed states. 

\subsection{Trial state implementation}
\subsubsection{Pure states}
We first consider the pure state case. The conventional way of realising the trial state is to apply a sequence of parameterised gates to an initial state as
$\ket{\phi(\vec{\theta})}  = R_{N}(\theta_N)\dots R_{k}(\theta_k)\dots R_{1}(\theta_1)\ket{\bar{0}}$. In practice, not only the gate, but also the measurement performed to the state can be also parametrised, which however can always be effectively realised  by applying a parameterised gate before a fixed measurement. Therefore, we also regard parametrised measurements equivalently as parametrised gates.
Next, we consider state preparation via post-selection of measurement outcomes. Specifically, after initialising the target system together with ancillae, we can apply a joint parametrised circuit, measure the ancillae, and post-select the trial state conditioned on the measurement outcome.  The quantum circuit of such a procedure is as follows. 
\begin{align*}
\Qcircuit @C=1em @R=.7em {
\lstick{\ket{\bar{0}}_A}&\multigate{1}{R(\vec{\theta})}
&\qw\\
\lstick{\ket{\bar{0}}_E}&\ghost{R(\vec{\theta})}&\meter\\
}
\end{align*}
As we only consider a pure trial state, the measurement should be a rank one projector, such as $\ket{0}\bra{0}_E$ without loss of generality. 
Therefore, the trial state is 
\begin{equation}\label{Eq:morecomplicated}
	\ket{\phi_0(\vec{\theta})}_A = \bra{0}_E\ket{\phi(\vec{\theta})}_{AE}/\sqrt{p(\vec{\theta})},
\end{equation}
with the joint state  before the measurement $\ket{\phi(\vec{\theta})}=R(\vec{\theta})\ket{\bar{0}}_{A}\ket{\bar{0}}_{E}$, and $p(\vec{\theta}) = |\bra{\phi(\vec{\theta})}_{AE}\ket{0}_E|^2$ being the post-selection probability of measuring $0$ of the ancillae.

\subsubsection{Mixed states and unitary evolution}
For mixed states under unitary evolution, we can design the trial states via two different ways. We can either input a mixed state $\rho(0)$ and apply a unitary circuit $R(\vec{\theta})$ to prepare the ansatz. \yx{Equivalently, we can input a state $\ket{\bar{0}}_{AE}$ of system $AE$ that is a purification of system $A$ satisfying  $\tr_E[\ket{\bar{0}}_{AE}\bra{\bar{0}}_{AE}] = \rho_A(0)$ with partial trace $\tr_E$ of system $E$.} \yx{Note that there is not a unique representation of the purified state and different choices may lead to greater or lower complexity of the simulation.} With the whole purified state, we only need to apply the circuit to system $A$ such as the following one.
\begin{align*}
\Qcircuit @C=1em @R=.7em {
&\gate{R(\vec{\theta})}&\qw\\
\lstick{\ket{\bar{0}}_{AE}}\\
&&\\
&\qw&\qw\\
}
\end{align*}
In this case, we use one trial state $\rhot = \tr_E[R(\vec{\theta})\ket{\bar{0}}_{AE}\bra{\bar{0}}_{AE}R^\dag(\vec{\theta})]$ with one parameter setting $\vec{\theta}$ to directly represent the state $\rho(t)$ at time $t$. Here, we can also consider more complicated ansatz with measurements such as Eq.~\eqref{Eq:morecomplicated}. 
Alternatively, we can decompose $\rho$ into pure states and simulate the unitary evolution of each pure state separately.
Suppose the initial state $\rho(0)$ has a decomposition $\rho(0)=\sum_ip_i\ket{\psi_i(0)}\bra{\psi_i(0)}$. In practice, one can  randomly input state $\ket{\psi_i(0)}$ with probability $p_i$ and evolve $\ket{\psi_i(t)}$ to time $t$ under the Hamiltonian $H$. Then the state $\rho(t)$ at time $t$ would be $\rho(t) = \sum_ip_i\ket{\psi_i(t)}\bra{\psi_i(t)}$. With VQS, $\ket{\psi_i(t)}$ is represented by the trial state $\ket{\phi(\vec{\theta}_i(t))}$ with parameters $\vec{\theta}_i(t)$. As the ansatz circuit $R(\vec{\theta})$ and parameters $\vec{\theta}$ can be different for the evolution of each pure state, evolving the pure states separately could lead to more accurate results than evolving the whole mixed state with one parameter settings. However, as the evolution equation for one trial state is obtained by minimising the distance of the whole state, evolving the pure states separately may not always lead more accurate simulation. 
Furthermore, it can also be practically hard to decompose the initial state into a mixture of pure states.


\subsubsection{Stochastic and imaginary time evolution}
For stochastic and imaginary time evolution evolutions, as they are generally not reversible, we have to introduce ancillae and jointly apply the parameterised circuit. We can also measure a subset of the ancillae to postselect the trial state. 
However, we can only use one trial state and it cannot be naively simulated by decomposing $\rho(0)$ into pure states as $\rho(0)=\sum_ip_i\ket{\psi_i(0)}\bra{\psi_i(0)}$ and evolving each pure state separately. This is because stochastic and imaginary time evolution is not a linear process acting on the initial state so that $\rho(T)\neq\sum_ip_i\rho_i(T)$ for real time $T=t$ and imaginary time $T=\tau$. Here $\rho(T)$ and $\rho_i(T)$ are the evolved state of $\rho(0)$ and $\ket{\psi_i(0)}$ at time $T$, respectively. 

In Ref.~\cite{suguru18theory}, we do present an alternative method to simulate the stochastic master equation with different pure states evolving with different parameters. While the simulation involves several new techniques including variational simulation of general unphysical processes, we refer the readers to Ref.~\cite{suguru18theory} for more information.


\subsection{Coefficients evaluation}
Now, we consider how to measure the coefficients $M$, $V$, and $Y$ in the evolution equations. 
For unitary evolution of mixed states, we can effective regard the input state as a pure state and only evolve the subsystem of $\rho_0$. Suppose the trial state is prepared by directly applying a unitary circuit to the initial state, the $M$, $V$, and $Y$ can all be measured with the circuit in Fig.~\ref{fig:circuit}. Instead, when the trial state is prepared conditioned on postselecting a measurement outcome as defined in Eq.~\eqref{Eq:morecomplicated}, we need to reconsider how to measure the coefficients. The derivative of the state can be written as
\begin{equation}\label{Eq:}
\begin{aligned}
	  	\frac{\partial\ket{\phi_0(\vec{\theta})}_A}{\partial\theta_k} = \frac{\bra{0}_E\frac{\partial\ket{\phit}_{AE}}{\partial \theta_k}}{\sqrt{p(\vec{\theta})}}-\frac{\bra{0}_E\ket{\phit}_{AE}\frac{\partial p(\vec{\theta})}{\partial \theta_k}}{2p^{3/2}(\vec{\theta})}
\end{aligned}
\end{equation}
The probability $p(\vec{\theta})$ can be directly measured and the derivative $\frac{\partial p(\vec{\theta})}{\partial \theta_k}$ can be measured by the circuit in Fig.~\ref{fig:circuit}
To evaluate $M$, $V$, and $Y$, we need to measure the following terms for all $i$ and $j$,
\begin{equation}\label{Eq:}
\begin{aligned}
	  		\Re\left(\frac{\partial \bra{\phi(\vec{\theta}(t))}}{\partial \theta_i}\frac{\partial \ket{\phi(\vec{\theta}(t))}}{\partial \theta_j}\right),&\,\Im\left(\frac{\partial \bra{\phi(\vec{\theta}(t))}}{\partial \theta_i}H\ket{\phi(\vec{\theta}(t))}\right),\\
	  		\Im\left(\frac{\partial \bra{\phi(\vec{\theta}(t))}}{\partial \theta_i}\ket{\phi(\vec{\theta}(t))}\right), &\,\Re\left(\frac{\partial \bra{\phi(\vec{\theta}(t))}}{\partial \theta_i}H\ket{\phi(\vec{\theta}(t))}\right).
\end{aligned}
\end{equation}
\yx{All these terms can be efficiently measured with the circuit shown in Fig.~\ref{fig:circuit} as they are in the form of elements of either $A$ or $C$ as defined in Eq.~\eqref{Eq:defAandC}. Note that the precision of the estimated derivatives depend on the size of the system or the number of measurements, which can affect the accuracy of the simulation. We refer to Ref.~\cite{Li2017} for a detailed analysis of the scaling in terms of the system size and the simulation accuracy.}


Now, we consider the case of general evolutions. Suppose the trial state is prepared by $\rho = \RR_{N} \RR_{N-1} \cdots \RR_1(\rho_0)$, where $\rho_0$ is a joint state of the system and ancillae and $\RR_k$ are quantum gates that applied on the whole system. 
We express the generator as
\begin{eqnarray}
\LL (\rho) = \sum_{i} g_{i} S_{i} \rho T_{i}^\dag,
\label{eq:Lin}
\end{eqnarray}
where $S_{i}$ and $T_{i}$ are unitary operators, and $g_{i}$ are coefficients. Similarly, we write
\begin{eqnarray}
\frac{\partial \RR_{k}(\rho)}{\partial \lambda_{k}} = \sum_{i} r_{k,i} S_{k,i} \rho T_{k,i}^\dag,
\label{eq:expansion2}
\end{eqnarray}
where $S_{k,i}$ and $T_{k,i}$ are unitary operators, and $r_{k,i}$ are coefficients. For example, consider gate $\RR_k(\theta_k)=e^{-i\theta_k\sigma/2}\rho e^{i\theta_k\sigma/2}$ with a one or two qubit Hermitian operator $\sigma$, the derivative is ${\partial \RR_{k}(\rho)}/{\partial \lambda_{k}} = -i\sigma/2\cdot e^{-i\theta_k\sigma/2}\rho e^{i\theta_k\sigma/2}+ e^{-i\theta_k\sigma/2}\rho e^{i\theta_k\sigma/2}\cdot i\sigma/2$. 

Using the expression (\ref{eq:expansion2}), we rewrite the derivative of the mixed state as
\begin{eqnarray}
\frac{\partial \rho}{\partial \lambda_{k}} = \sum_{i} r_{k,i} \RR_{k,i} \rho_0,
\end{eqnarray}
where
\begin{eqnarray}
\RR_{k,i} \rho_0 &=& \RR_{N} \RR_{N-1} \cdots  \RR_{k+1} [ S_{k,i} (\RR_{k-1} \cdots \RR_{2} \RR_{1} \rho_0) T_{k,i}^\dag].
\end{eqnarray}
Then, using the expression (\ref{eq:Lin}), coefficients can be expressed as
\begin{equation}\label{Eq:MVY}
\begin{aligned}
	  {M}_{k,q}&=\sum_{i,j} \frac{1}{2} \{ r_{k,i}r_{q,j} \Tr [
(\RR_{k,i} \rho_0) (\RR_{q,j} \rho_0)
] + \HC \},\\
V_k&= \sum_{i,j}  \frac{1}{2} \{ r_{k,i} g_{j} \Tr [
(\RR_{k,i} \rho_0) (S_{j} \rho T_{j}^\dag)
]  + \HC \},\\
Y_k&= -\sum_{i}  \frac{1}{2} \{ r_{k,i} \Tr [
(\RR_{k,i} \rho_0) \rho H)
]  + \HC \}
\end{aligned}
\end{equation}

\begin{figure*}[t]
\centering
\includegraphics[width=0.75\linewidth]{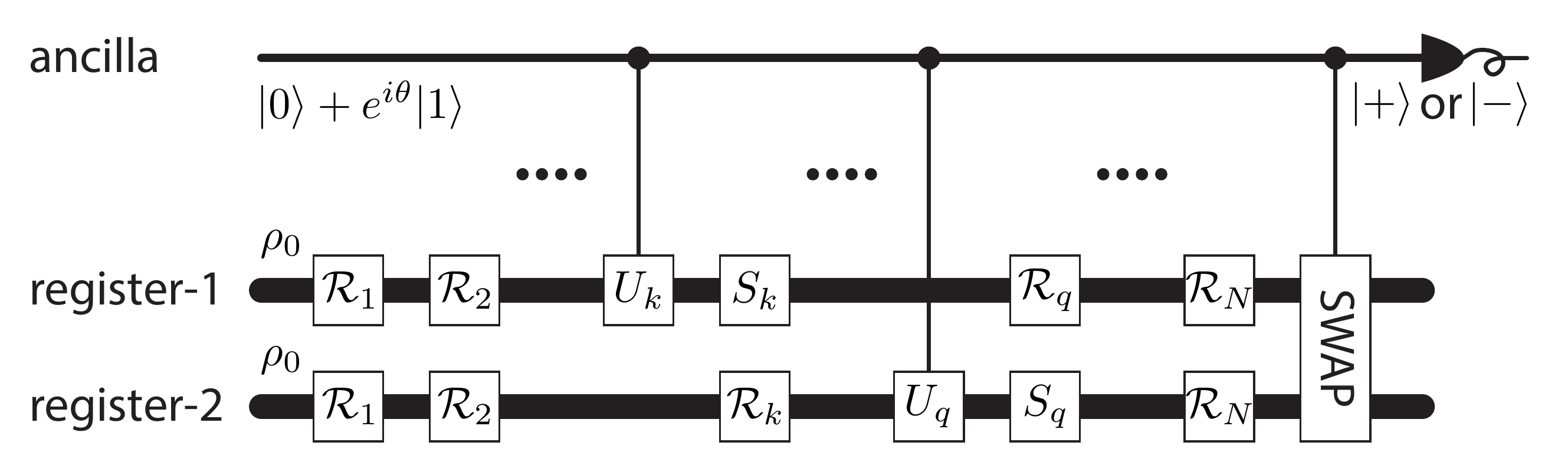}
\caption{
Circuit for the evaluation of differential-equation coefficients of the mixed-state simulator. To evaluate $\Re \left[ e^{i\theta} \Tr (\rho_1 \rho_2) \right]$, where $\rho_1 = \RR_{N} \RR_{N-1} \cdots \RR_{k+1} [ S_{k} (\RR_{k-1} \cdots \RR_{2} \RR_{1} \rho_0) T_{k}^\dag]$, $\rho_2 = \RR_{N} \RR_{N-1} \cdots \RR_{q+1} [ S_{q} (\RR_{q-1} \cdots \RR_{2} \RR_{1} \rho_0) T_{q}^\dag]$, the ancillary qubit is initialised in the state $(\ket{0}+e^{i\theta}\ket{1})/\sqrt{2}$ and measured in the $\ket{\pm} = (\ket{0} \pm \ket{1})/\sqrt{2}$ basis. Here, $U_{k} = T_{k}S_{k}^\dag$ and $U_{q} = T_{q}S_{q}^\dag$. The last gate is a controlled swap gate on two registers. In the figure, we have assumed that $k < q$, and $q = N+1$ when the circuit is used to evaluate $\tilde{V}_{k}$ coefficients.
}
\label{fig:circuit2}
\end{figure*}

In Eq.~\eqref{Eq:MVY}, each term is in the from
$$
a \Re \left[ e^{i\theta} \Tr (\rho_1 \rho_2) \right],
$$
where the amplitude $a$ and phase $\theta$ are determined by either $r_{k,i}r_{q,j}$ or $r_{k,i} g_{j}$. We would like to remark that, in general,  $\rho_1$ and $\rho_2$ are not reduced density matrices as $S$ and $T$ are separately applied to the left and right sides of the state. Nevertheless, such a term can be evaluated with the quantum circuit shown in Fig.~\ref{fig:circuit2}. This circuit needs an ancillary qubit initialised in the state $(\ket{0}+e^{i\theta}\ket{1})/\sqrt{2}$ and two registers initialised in the state $\rho_0$. The ancillary qubit is measured after a sequence of operations on registers and some controlled unitary operations, where the ancillary qubit is the control qubit. The value is given by $\Re [ e^{i\theta} \Tr (\rho_1 \rho_2) ] = 2P_{+} - 1$, where $P_{+}$ is the probability that the qubit is in the state $\ket{+} = (\ket{0}+\ket{1})/\sqrt{2}$. We can also consider trial state prepared by post-selection and the matrix elements can be similarly evaluated.

\subsection{Numerical simulation}
\yx{Here, we show a numerical example of the variational algorithm for simulating the real dynamics of open quantum systems. We consider a two-qubit Ising model under independent amplitude damping noise with the Lindblad master equation
\begin{equation}\label{eq:Lindbladexample}
\begin{aligned}
\frac{d}{dt}\rho &=-i[H, \rho]+\sum_{j=0}^1 \mathcal{L}_j(\rho).
\end{aligned}
\end{equation}
Here the system Hamiltonian is $H=X_0 +X_1 + \frac{1}{4}Z_0 Z_1$, the Lindblad terms are $\mathcal{L}_j(\rho)=\frac{1}{2}(2\sigma^-_j \rho \sigma^+_j-\sigma^+_j \sigma^-_j \rho -\rho \sigma^+_j \sigma^-_j )$, and $\sigma ^- _j=\ket{0}\bra{1}_j$ and $\sigma ^+ _j=\ket{1}\bra{0}_j$ are lowering and raising operators acting on the $j^{\textrm{th}}$  qubit, respectively. 
We start with a product initial state $\rho_0 = \ket{00}\bra{00}$ and simulate the dynamics from $t=0$ to $t=10$. }

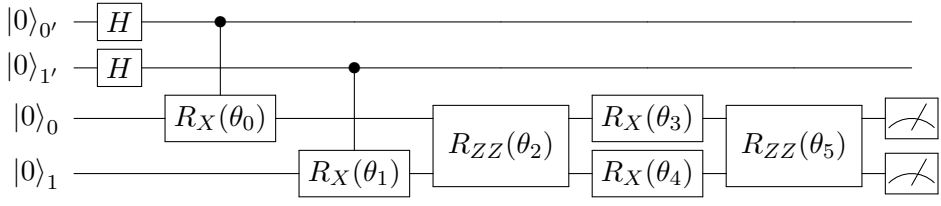
\begin{figure}[t]
\begin{flushright}
\begin{align*}
\Qcircuit @C=.8em @R=.3em {
\lstick{\ket{0}_{0'}}&\gate{H}&\ctrl{2} &\qw &\qw & \qw &\qw &\qw \\
\lstick{\ket{0}_{1'}}&\gate{H}&\qw &\ctrl{2} & \qw & \qw & \qw &\qw \\
\lstick{\ket{0}_0}& \qw & \gate{R_X(\theta_0)} &\qw &\multigate{1}{R_{ZZ}(\theta_2)}& \gate{R_X(\theta_3)} &\multigate{1}{R_{ZZ}(\theta_5)} &\meter \\
\lstick{\ket{0}_1}& \qw & \qw & \gate{R_X(\theta_1)}&\ghost{R_{ZZ}(\theta_5)}& \gate{R_X(\theta_4)}&\ghost{R_{ZZ}(\theta_5)} & \meter
}
\end{align*}
\end{flushright}
\caption{The ansatz used in our numerical simulation. We introduced two ancillary qubits $0'1'$ as a purification of the two system qubits $01$.  We define $R_X(\theta_i) = e^{-i\theta_iX}$ and $R_{ZZ}(\theta) = e^{-i \theta_i Z_1 \otimes Z_2}$, which are the evolution of the system Hamiltonian terms. We used six parameters in total. }  
\label{Fig:ansatz}
\end{figure}

\yx{For the variational algorithm, we consider the Hamiltonian ansatz as shown in Fig.~\ref{Fig:ansatz}, where we introduced two ancillary qubits to purify the state. For the ansatz, we first apply parametrised gates to entangle   the ancillae and the system qubits; then we apply the Hamiltonian ansatz, which consists of parameterised gates determined by the Hamiltonian of the system, to the system qubits. In general, the ansatz can also depend on the Lindblad terms and we leave the design of ansatz for general open system evolutions in a future work.  }

\yx{To simulate the evolution of Eq.~\eqref{eq:Lindbladexample}, we consider discretised timestep $\delta t=0.01$. For the initial step, we set all the parameters to $0$; for the later steps, we evaluate all terms of $M$ and $V$ defined in Eq.~(\ref{eq:MVmatrixFirst})  and solve Eq.~(\ref{Eq:MVequation}) to update the parameters. }

\begin{figure}[t]
\centering
\includegraphics[width=0.6\linewidth]{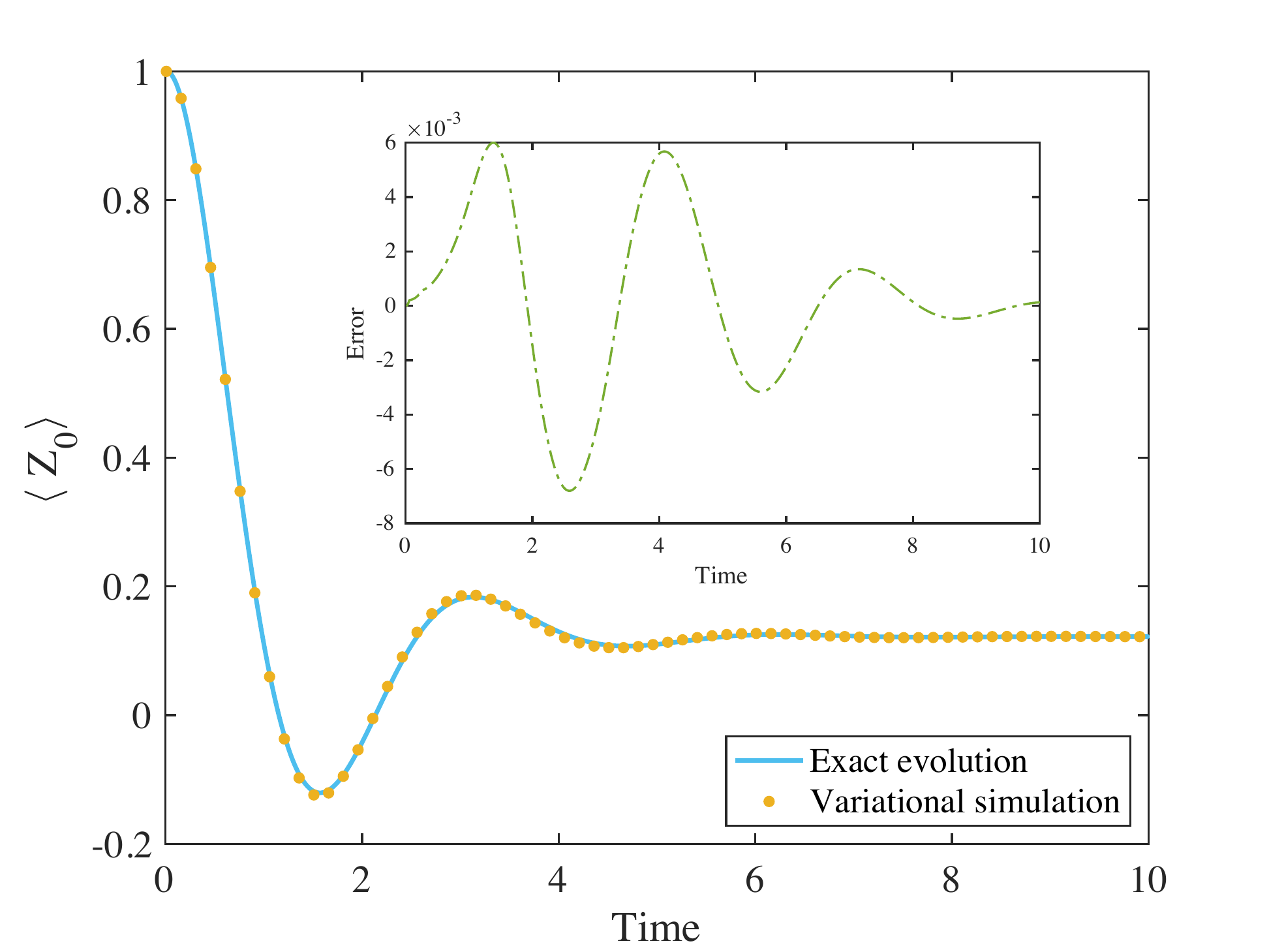}
\caption{The comparison between the exact result and the result obtained from variational algorithm. We set the time step $\delta t=0.01$ and simulated from $t=0$ to $t=10$. We used a software package QuTIP for the numerical simulation~\cite{johansson2012qutip,johansson2013qutip}. }
\label{fig:result}
\end{figure}

\yx{The simulation result is shown in Fig.~\ref{fig:result}. We also numerically solve the exact evolution of Eq.~(\ref{eq:Lindbladexample})  and compare it with the result obtained from our variational algorithm. We measure the expectation value of $Z_0$ and we found excellent agreement between the results from the exact solution and our variational algorithm with a deviation less than $10^{-2}$. }


\section{Discussion}\label{Sec:Discussion}
In this work, we focus on the theory of variational quantum simulation. We first study the equivalence and difference of the three variational principles. 
We find McLachlan's variational principle is the most consistent principle for variational quantum simulation with quantum gates controlled by real parameters. 
Then, we extend variational quantum simulation of pure states to the general evolution of mixed states under both real and imaginary time. We discuss possible realisation of the trial states and show how to efficiently implement the simulation with quantum circuits.  

In future works, one can study the design of trials states for specific problems and test our theory for simulating open quantum systems and preparing Gibbs states. It is  also interesting to experimentally realise our variational simulation algorithms with current and near-term noisy quantum hardware. As the variational method only uses shallow quantum circuits, error mitigation techniques can be applied to suppress errors \cite{subspace1,Li2017,PhysRevLett.119.180509, endo2017practical, subspace2,recoveringnoisefree, samerrormitigation,bonet2018low,mcclean2019decoding}. 

\section*{Acknowledgements.}
This work is supported by the EPSRC National Quantum Technology Hub in Networked Quantum Information Technology (EP/M013243/1).
SE is supported by Japan Student Services Organization (JASSO) Student Exchange Support Program (Graduate Scholarship for Degree Seeking
Students). QZ acknowledges support by the National Natural Science Foundation of China Grant No. 11674193. 
YL is supported by NSAF (Grant No. U1730449).

\onecolumn\newpage
\appendix

\section*{Appendix}

\section{Real time evolution}
\subsection{Real time evolution: pure state and unitary evolution}
\subsubsection{The Dirac and Frenkel variational principle}
The Schr\"odinger equation is,
\begin{equation}\label{SUP:Schrodinger}
	\frac{d \ket{\psi(t)}}{d t} =-i H\ket{\psi(t)}.
\end{equation}
Consider a parametrised trial state $\ket{\phi(\vec{\theta}(t))}$, with $\vec{\theta}(t) = (\theta_1(t), \theta_2(t),\dots, \theta_N(t))$, the real time evolution of the Schr\"odinger equation on the trial state space is
\begin{equation}\label{diffEq}
	\frac{d \ket{\phi(\vec{\theta}(t))}}{d t}  = \sum_i \frac{\partial \ket{\phi(\vec{\theta}(t))}}{\partial \theta_i}\dot{\theta_i}=-iH\ket{\phi(\vec{\theta}(t))}.
\end{equation}
Such an equation has a state vector on both sides. While, state $\frac{\partial \ket{\phi(\vec{\theta}(t))}}{\partial \theta_i}$ can be regarded as the tangent vector of state $\ket{\phi(\vec{\theta}(t))}$ at $\vec{\theta}(t)$, we can thus apply a projector 
\begin{equation}
	P = \sum_i\frac{\partial \ket{\phi(\vec{\theta}(t))}}{\partial \theta_i}\frac{\partial \bra{\phi(\vec{\theta}(t))}}{\partial \theta_i}
\end{equation}
to project the right hand side onto the tangent space. Therefore, the equation becomes,
\begin{equation}
	\begin{aligned}
		P\sum_j \frac{\partial \ket{\phi(\vec{\theta}(t))}}{\partial \theta_j}\dot{\theta}_j&=-iPH\ket{\phi(\vec{\theta}(t))},\\
	i\sum_i\frac{\partial \ket{\phi(\vec{\theta}(t))}}{\partial \theta_i}\frac{\partial \bra{\phi(\vec{\theta}(t))}}{\partial \theta_i}\sum_j \frac{\partial \ket{\phi(\vec{\theta}(t))}}{\partial \theta_j}\dot{\theta}_j&=-i\sum_i\frac{\partial \ket{\phi(\vec{\theta}(t))}}{\partial \theta_i}\frac{\partial \bra{\phi(\vec{\theta}(t))}}{\partial \theta_i}H\ket{\phi(\vec{\theta}(t))}.
	\end{aligned}
\end{equation}
For each term of $\frac{\partial \ket{\phi(\vec{\theta}(t))}}{\partial \theta_i}$, the coefficient should satisfy
\begin{equation}
	\sum_j \frac{\partial \bra{\phi(\vec{\theta}(t))}}{\partial \theta_i}\frac{\partial \ket{\phi(\vec{\theta}(t))}}{\partial \theta_j}\dot{\theta}_j=-i\frac{\partial \bra{\phi(\vec{\theta}(t))}}{\partial \theta_i}H\ket{\phi(\vec{\theta}(t))}.
\end{equation}
Define the matrix elements of $M$ and $V$ as
\begin{equation}
	\begin{aligned}
		A_{i,j} &= \frac{\partial \bra{\phi(\vec{\theta}(t))}}{\partial \theta_i}\frac{\partial \ket{\phi(\vec{\theta}(t))}}{\partial \theta_j},\\
		C_i &= \frac{\partial \bra{\phi(\vec{\theta}(t))}}{\partial \theta_i}H\ket{\phi(\vec{\theta}(t))},
	\end{aligned}
\end{equation}
the evolution of parameters is simplified to
\begin{equation}
	\sum_j A_{i,j}\dot{\theta}_j = -iC_i.
\end{equation}

The same equation can be obtained by applying the Dirac and Frenkel variational principle 
\begin{equation}
	\left\langle{ \delta\phit\bigg|\left(\frac{d}{d t}+iH\right)\bigg|\phi(\vec{\theta}(t))}\right\rangle = 0.
\end{equation}
This is can be verified with 
\begin{equation}
	\bra{\delta \phit} = \sum_i \frac{\partial \bra{\phi(\vec{\theta}(t))}}{\partial \theta_i}\delta{\theta_i},
\end{equation}
and Eq.~\eqref{diffEq}.

\subsubsection{McLachlan's variational principle}
The McLachlan's variational principle \cite{McLachlan}, applied to real time evolution, is given by
 \begin{equation}
 	\delta \|({d}/{d t} +iH)\ket{\phit}\|=0
 \end{equation}
where
\begin{equation}
\begin{aligned}
	\|({d}/{d t} +iH)\ket{\phit}\|^2=&\left({d}/{d t} +iH)\ket{\psi(t)}\right)^\dag({d}/{d t} +iH)\ket{\phit},\\
	=&\sum_{i,j}\frac{\partial \bra{\phit}}{\partial \theta_i}\frac{\partial \ket{\phit}}{\partial \theta_j}\dot{\theta}_i^* \dot{\theta}_j+ i\sum_{i}\frac{\partial \bra{\phit}}{\partial \theta_i}H\ket{\phit}\dot{\theta}_i^*\\
		&- i\sum_{i}\bra{\phit}H\frac{\partial \ket{\phit}}{\partial \theta_i}\dot{\theta}_i +\bra{\phit}	H^2\ket{\phit}. 
\end{aligned}
\end{equation}

Note that the variation respect to $\|({d}/{d t} +iH)\ket{\phit}\|$ is equivalent to the variation respect to $\|({d}/{d t} +iH)\ket{\phit}\|^2$, so we focus on $\|({d}/{d t} +iH)\ket{\phit}\|^2$ instead. 

Suppose $\dot{\theta}_i$ can be complex, then  we have
\begin{equation}
	\begin{aligned}
		\delta \|({d}/{d t} + iH)\ket{\phit}\|^2=& \left(\sum_{j}\frac{\partial \bra{\phit}}{\partial \theta_i}\frac{\partial \ket{\phit}}{\partial \theta_j} \dot{\theta}_j+ i\frac{\partial \bra{\phit}}{\partial \theta_i}H\ket{\phit}\right)\delta\dot{\theta}_i^*,\\
		&+\left(\sum_{j}\frac{\partial \bra{\phit}}{\partial \theta_j}\frac{\partial \ket{\phit}}{\partial \theta_i} \dot{\theta}_j^*- i\frac{\partial \bra{\phit}}{\partial \theta_i}H\ket{\phit}\right)\delta\dot{\theta}_i.
\end{aligned}
\end{equation}
Then the evolution is 
\begin{equation}
	\sum_{j}\frac{\partial \bra{\phit}}{\partial \theta_i}\frac{\partial \ket{\phit}}{\partial \theta_j} \dot{\theta}_j= -i\frac{\partial \bra{\phit}}{\partial \theta_i}H\ket{\phit}.
\end{equation}

On the other hand, suppose $\dot{\theta}_i$ is real, then 
\begin{equation}
	\begin{aligned}
		\delta \|({d}/{d t} + iH)\ket{\phit}\|^2=&\sum_{j}\left(\frac{\partial \bra{\phit}}{\partial \theta_i}\frac{\partial \ket{\phit}}{\partial \theta_j} + \frac{\partial \bra{\phit}}{\partial \theta_j}\frac{\partial \ket{\phit}}{\partial \theta_i}\right) \dot{\theta}_j\delta\theta_i\\
		&+i\left(\frac{\partial \bra{\phit}}{\partial \theta_i}H\ket{\phit}-\bra{\phit}H\frac{\partial \ket{\phit}}{\partial \theta_i}\right)\delta\theta_i.
\end{aligned}
\end{equation}
The corresponding evolution equation for parameters is
\begin{equation}
	\begin{aligned}
		&\sum_{j}\left(\frac{\partial \bra{\phit}}{\partial \theta_i}\frac{\partial \ket{\phit}}{\partial \theta_j} + \frac{\partial \bra{\phit}}{\partial \theta_j}\frac{\partial \ket{\phit}}{\partial \theta_i}\right) \dot{\theta}_j\\
		=&-i\left(\frac{\partial \bra{\phit}}{\partial \theta_i}H\ket{\phit}-\bra{\phit}H\frac{\partial \ket{\phit}}{\partial \theta_i}\right)
	\end{aligned}
\end{equation}
or equivalently
\begin{equation}
	\sum_j A^R_{i,j}\dot{\theta}_j = C_i^I,
\end{equation}
where $A^R_{i,j}$ and $C_i^I$ are the real and imaginary parts of $A_{i,j}$ and $C_i$, respectively.

\subsubsection{Time-dependent variational principles}
The Lagrangian for the Schr\"odinger equation is
\begin{equation}
\begin{aligned}
	L &= \left\langle{{\psi}\bigg|\left(\frac{d}{d t}+iH\right)}\bigg|{\psi}\right\rangle,
\end{aligned}
\end{equation}
and the Schr\"odinger equation is obtained by 
\begin{equation}
	\frac{\partial L}{\partial \bra{\psi}} - \frac{d}{dt}\frac{\partial L}{\partial \frac{\partial\bra{\psi}}{\partial t}} = \left(\frac{d}{d t}+iH\right)\ket{\psi} = 0.
\end{equation}
Suppose $\ket{\psi} = \ket{\phit}$, then the Lagrangian is 
\begin{equation}
\begin{aligned}
	L &= \left\langle{{\phit}\bigg|\left(\frac{d}{d t}+iH\right)}\bigg|{\phit}\right\rangle,\\
	&=\sum_i\dot{\theta_i}\left\langle{{\phit}\bigg|\frac{\partial}{\partial \theta_i}}\bigg|{\phit}\right\rangle +i\left\langle{{\phit}\bigg|H}\bigg|{\phit}\right\rangle
\end{aligned}
\end{equation}
Suppose $\theta$ is complex, then 
\begin{equation}
\begin{aligned}
	\frac{\partial L}{\partial \theta_i^*} - \frac{d}{dt}\frac{\partial L}{\partial \dot\theta_i^*} = \sum_j\dot{\theta_j}\frac{\partial\bra{\phit}}{\partial \theta_i}\frac{\partial\ket{\phit}}{\partial \theta_j}+i\frac{\partial \bra{\phit}}{\partial \theta_i}H\ket{\phit},
\end{aligned} 
\end{equation}
and the evolution is
\begin{equation}
	\sum_j\frac{\partial\bra{\phit}}{\partial \theta_i}\frac{\partial\ket{\phit}}{\partial \theta_j}\dot{\theta_j} = -i\frac{\partial \bra{\phit}}{\partial \theta_i}H\ket{\phit}. 
\end{equation}

Instead, suppose $\theta$ is real, then we have
\begin{equation}
\begin{aligned}
	\frac{\partial L}{\partial \theta_i} - \frac{d}{dt}\frac{\partial L}{\partial \dot\theta_i} =& \sum_j\dot{\theta}_j\frac{\partial \left\langle{{\phit}\bigg|\frac{\partial}{\partial \theta_j}}\bigg|{\phit}\right\rangle}{\partial \theta_i} +i \frac{\partial \left\langle{{\phit}\bigg|H}\bigg|{\phit}\right\rangle}{\partial \theta_i} - \frac{d}{dt}\left\langle{{\phit}\bigg|\frac{\partial}{\partial \theta_i}}\bigg|{\phit}\right\rangle,\\
	=&\sum_j\dot{\theta}_j\left(\frac{\partial\bra{\phit}}{\partial \theta_i}\frac{\partial\ket{\phit}}{\partial \theta_j}+\left\langle{{\phit}\bigg|\frac{\partial^2}{\partial \theta_i\partial \theta_j}}\bigg|{\phit}\right\rangle\right) +i\frac{\partial \bra{\phit}}{\partial \theta_i}H\ket{\phit}\\
	&+i\bra{\phit}H\frac{\partial \ket{\phit}}{\partial \theta_i}-
	\sum_{j} \dot{\theta}_j\left(\frac{\partial\bra{\phit}}{\partial \theta_j}\frac{\partial\ket{\phit}}{\partial \theta_i} 
	+\left\langle{{\phit}\bigg|\frac{\partial^2}{\partial \theta_i\partial \theta_j}}\bigg|{\phit}\right\rangle\right),\\
	=&\sum_j\dot{\theta}_j\left(\frac{\partial\bra{\phit}}{\partial \theta_i}\frac{\partial\ket{\phit}}{\partial \theta_j}-\frac{\partial\bra{\phit}}{\partial \theta_j}\frac{\partial\ket{\phit}}{\partial \theta_i}
\right) \\
&+i\frac{\partial \bra{\phit}}{\partial \theta_i}H\ket{\phit}+i\bra{\phit}H\frac{\partial \ket{\phit}}{\partial \theta_i},
\end{aligned} 
\end{equation}
and
\begin{equation}
	\begin{aligned}
		&\sum_j\left(\frac{\partial\bra{\phit}}{\partial \theta_i}\frac{\partial\ket{\phit}}{\partial \theta_j}-\frac{\partial\bra{\phit}}{\partial \theta_j}\frac{\partial\ket{\phit}}{\partial \theta_i}
\right)\dot{\theta}_j\\
 =&-i\left(\frac{\partial \bra{\phit}}{\partial \theta_i}H\ket{\phit}+\bra{\phit}H\frac{\partial \ket{\phit}}{\partial \theta_i}\right),
	\end{aligned}
\end{equation}
Equivalently, the evolution to parameters is
\begin{equation}
		\sum_j A_{i,j}^I \dot\theta_j = -C_i^R,
\end{equation}
where, $A^I_{i,j}$ and $C_i^R$ are the imaginary and real parts of $A_{i,j}$ and $C_i$, respectively.

\subsection{Mixed states and general evolution}\label{Sec:appendix2}
\subsubsection{General evolution}
Suppose the initial state is a mixed state $\rho$, the stochastic evolution is defined by
\begin{equation}
	\frac{d \rho}{d t} = \ml. 
\end{equation}
Consider a parameterised trial state $\rhot$, we can get the evolution of $\theta$ that simulates the stochastic evolution.

\paragraph{The time-dependent variational principle.}
Let $\rho = \rhot$, the stochastic evolution equation is
\begin{equation}
	\sum_i \frac{\partial \rhot}{\partial \theta_i}\dot{\theta}_i = \ml. 
\end{equation}
Because $\frac{d \rhot}{d t} = \sum_i \frac{\partial \rhot}{\partial \theta_i}\dot{\theta}_i$ is in the tangent subspace of $\{\frac{\partial \rhot}{\partial \theta_i}\}$, we need to project the stochastic evolution  equation onto this tangent subspace. Following the Dirac and Frenkel variational principle, we have
\begin{equation}
	\tr\left[(\delta\rhot)^\dag\left(\frac{d \rho}{d t}-\ml\right)\right] = \tr\left[\left(\sum_i\frac{\partial \rhot}{\partial \theta_i}\delta\theta_i\right)^\dag\left(\sum_j\frac{\partial \rho}{\partial \theta_j}\dot\theta_j-\ml\right)\right] = 0.
\end{equation}
The evolution of parameters is
\begin{equation}
M_{i,j}\dot{\theta}_j=V_i,
\end{equation}
with 
\begin{equation}
	\begin{aligned}
		M_{i,j} &= \tr\left[\left(\frac{\partial \rhot}{\partial \theta_i}\right)^\dag\frac{\partial \rhot}{\partial \theta_j}\right],\\
		V_i &= \tr\left[\left(\frac{\partial \rhot}{\partial \theta_i}\right)^\dag\ml\right].
	\end{aligned}
\end{equation}
Note that when parameters $\theta$ are real, $M$ and $V$ are also real and the solution $\dot{\theta}_j$ is also real. 

\paragraph{McLachlan's variational principle.}
We can also minimise the distance between the evolution via parameters and the stochastic evolution via McLachlan's variational principle,
\begin{equation}
 	\delta \|{d \rho}/{d t}-\ml\|=0,
\end{equation}
where 
\begin{equation}
\begin{aligned}
	\|{d \rho}/{d t} -\ml\|^2=&\tr\left[({d \rho}/{d t} -\ml)^\dag({d \rho}/{d t} -\ml)\right],\\
	=&\tr\left[\left(\sum_i\frac{\partial \rhot}{\partial \theta_i}\dot\theta_i\right)^\dag\sum_j\frac{\partial \rhot}{\partial \theta_j}\dot\theta_j-\left(\sum_i\frac{\partial \rhot}{\partial \theta_i}\dot\theta_i\right)^\dag\ml\right]\\
	&+\tr\left[-\left(\sum_i\frac{\partial \rhot}{\partial \theta_i}\dot\theta_i\right)\ml+\ml^2\right].
\end{aligned}
\end{equation}

Note that the variation with respect to $\|{d \rho}/{d t} -\ml\|$ is equivalent to the variation with respect to $\|{d \rho}/{d t} -\ml\|^2$, so we focus on $\|{d \rho}/{d t} -\ml\|^2$ instead. 

Suppose $\dot \theta$ is complex, then
\begin{equation}
\begin{aligned}
	\delta\|{d \rho}/{d t} -\ml\|^2=&\tr\left[\left(\frac{\partial \rhot}{\partial \theta_i}\right)^\dag\sum_j\frac{\partial \rhot}{\partial \theta_j}\dot\theta_j-\left(\frac{\partial \rhot}{\partial \theta_i}\right)^\dag\ml\right]\delta\dot\theta_i^*\\
	&+\tr\left[\left(\sum_j\frac{\partial \rhot}{\partial \theta_j}\dot\theta_j\right)^\dag\frac{\partial \rhot}{\partial \theta_i}-\frac{\partial \rhot}{\partial \theta_i}\ml\right]\delta\dot\theta_i.
\end{aligned}
\end{equation}
The evolution to $\dot \theta$ is
\begin{equation}
	\sum_j\tr\left[\left(\frac{\partial \rhot}{\partial \theta_i}\right)^\dag\frac{\partial \rhot}{\partial \theta_j}\right]\dot\theta_i=\tr\left[\left(\frac{\partial \rhot}{\partial \theta_i}\right)^\dag\ml\right]
\end{equation}

When $\dot \theta$ is real, then 
\begin{equation}
\begin{aligned}
	\delta\|{d \rho}/{d t} -\ml\|^2=&\tr\left[\left(\frac{\partial \rhot}{\partial \theta_i}\right)\sum_j\frac{\partial \rhot}{\partial \theta_j}\dot\theta_j-\frac{\partial \rhot}{\partial \theta_i}\ml\right]\delta\dot\theta_i\\
	&+\tr\left[\left(\sum_j\frac{\partial \rhot}{\partial \theta_j}\dot\theta_j\right)\frac{\partial \rhot}{\partial \theta_i}-\frac{\partial \rhot}{\partial \theta_i}\ml\right]\delta\dot\theta_i.
\end{aligned}
\end{equation}
The evolution to $\dot \theta$ is
\begin{equation}
	\sum_j\tr\left[\frac{\partial \rhot}{\partial \theta_i}\frac{\partial \rhot}{\partial \theta_j}\right]\dot\theta_j=\tr\left[\frac{\partial \rhot}{\partial \theta_i}\ml\right]
\end{equation}

Then we can get the same equation for parameters
\begin{equation}
M_{i,j}\dot{\theta}_j=V_i.
\end{equation} 

\paragraph{The time-dependent variational principle.}
The Lagrangian for a general stochastic evolution is
\begin{equation}
	L = \tr[\rho^\dag({d \rho}/{d t} -\ml)]. 
\end{equation}
the Schr\"odinger equation is obtained by 
\begin{equation}
	\frac{\partial L}{\partial \rho^\dag} - \frac{d}{dt}\frac{\partial L}{\partial \frac{\partial\rho^\dag}{\partial t}} = {d \rho}/{d t} -\ml = 0.
\end{equation}
Suppose $\rho = \rhot$, then the Lagrangian is 
\begin{equation}
\begin{aligned}
	L &= \tr[\rhot^\dag({d \rhot}/{d t}  -\ml)],\\
	&=\tr\left[\rhot^\dag\left(\sum_j\frac{\partial \rhot}{\partial \theta_j}\dot\theta_j -\ml\right)\right]
	\end{aligned}
\end{equation}
Suppose $\theta$ is complex, then 
\begin{equation}
\begin{aligned}
	\frac{\partial L}{\partial \theta_i^*} - \frac{d}{dt}\frac{\partial L}{\partial \dot\theta_i^*} = \tr\left[\frac{\partial \rhot^\dag}{\partial \theta_i^*}\left(\sum_j\frac{\partial \rhot}{\partial \theta_j}\dot\theta_j -\ml\right)\right],
\end{aligned} 
\end{equation}
and the evolution is
\begin{equation}
	\sum_j\tr\left[\left(\frac{\partial \rhot}{\partial \theta_i}\right)^\dag\frac{\partial \rhot}{\partial \theta_j}\right]\dot\theta_j=\tr\left[\left(\frac{\partial \rhot}{\partial \theta_i}\right)^\dag\ml\right]
\end{equation}

Instead, suppose $\theta$ is real, we cannot reproduce the stochastic evolution equation and hence cannot obtain the evolution equation of the parameters. 

\subsubsection{Reduction to the pure state case}
Consider unitary evolution of mixed input states, the von Neumann equation has $\mathcal{L}(\rho) = -i[H, \rho(t)]$,
\begin{equation}
	\frac{d \rho(t)}{d t} = -i[H, \rho(t)]. 
\end{equation}
In this case, $M$  and $V$ are explicitly given by
\begin{equation}
	\begin{aligned}
	M_{i,j} &= \tr\left[\left(\frac{\partial \rhot}{\partial \theta_i}\right)^\dag\frac{\partial \rhot}{\partial \theta_j}\right],\\
		V_i &= -i\tr\left[\left(\frac{\partial \rhot}{\partial \theta_i}\right)^\dag[H, \rhot]\right].
	\end{aligned}
\end{equation}

Suppose $\rhot$ is a pure state $\ket{\phit}\bra{\phit}$, the von Neumann equation becomes the Schr\"odinger equation. For the variational evolution equation, we have
\begin{equation}
	\begin{aligned}
		M_{i,j} &= \tr\left[\frac{\partial \rhot}{\partial \theta_i}\frac{\partial \rhot}{\partial \theta_j}\right],\\
		 &= \tr\left[\left(\frac{\partial \ket{\phit}}{\partial \theta_i}\bra{\phit}+\ket{\phit}\frac{\partial \bra{\phit}}{\partial \theta_i}\right)\left(\frac{\partial \ket{\phit}}{\partial \theta_j}\bra{\phit}+\ket{\phit}\frac{\partial \bra{\phit}}{\partial \theta_j}\right)\right],\\
		&= \Re\left(\frac{\partial \bra{\phit}}{\partial \theta_i}\frac{\partial \ket{\phit}}{\partial \theta_j}+\frac{\partial \bra{\phi(\vec{\theta}(t))}}{\partial \theta_i}\ket{\phit}\frac{\partial \bra{\phi(\vec{\theta}(t))}}{\partial \theta_j}\ket {\phit}\right) ,\\ 
		&= A_{i,j}^R+ \frac{\partial \bra{\phi(\vec{\theta}(t))}}{\partial \theta_i}\ket{\phit}\frac{\partial \bra{\phi(\vec{\theta}(t))}}{\partial \theta_j}\ket {\phit},\\
		V_i &= -i\tr\left[\frac{\partial \rhot}{\partial \theta_i}[H, \rhot]\right],\\
		&= -i\tr\left[\left(\frac{\partial \ket{\phit}}{\partial \theta_i}\bra{\phit}+\ket{\phit}\frac{\partial \bra{\phit}}{\partial \theta_i}\right)\left(H\ket{\phit}\bra{\phit} - \ket{\phit}\bra{\phit} H\right)\right],\\
		&=\Im\left(\frac{\partial \bra{\phit}}{\partial \theta_i}H\ket{\phit}+\bra{\phit}\frac{\partial \ket{\phi(\vec{\theta}(t))}}{\partial \theta_i}\bra{\phit}H\ket{\phi(\vec{\theta}(t))}\right),\\
		& = C_i^I+i\frac{\partial \bra{\phi(\vec{\theta}(t))}}{\partial \theta_i}\ket{\phit}\bra{\phit}H\ket{\phi(\vec{\theta}(t))}
	\end{aligned} 
\end{equation}
This equation is equivalent to applying a phase gate to an additional ancilla with an additional parameter in the pure state equation.
\begin{proof}
	Suppose there are $N$ parameters in the ansatz. Consider an additional ancilla that is rotated via a phase gate $e^{i\theta_{0}}$, the evolution of the $N+1$ parameters via the pure state equation is equivalent to the evolution of the $N$ parameters via the mixed state equation. Under the pure state equation, the evolution of the $N+1$ parameters is
	\begin{equation}
		\begin{aligned}
			\sum_{j=1}^N A^R_{i,j}\dot\theta_j + i\frac{\partial \bra{\phi(\vec{\theta}(t))}}{\partial \theta_i}\ket{\phit}\dot\theta_{0}&= C_i^I,\\
			-i\sum_{j=1}^N \frac{\partial \bra{\phi(\vec{\theta}(t))}}{\partial \theta_i}\ket{\phit}\dot\theta_j-\dot\theta_{0}&= \bra{\phit}H\ket{\phi(\vec{\theta}(t))}.
		\end{aligned}
	\end{equation}
	Substituting the last equation into the first $N$ equations, we get
		\begin{equation}
		\begin{aligned}
			&\sum_{j=1}^N A^R_{i,j}\dot\theta_j + i\frac{\partial \bra{\phi(\vec{\theta}(t))}}{\partial \theta_i}\ket{\phit}\left(-i\sum_{j=1}^N \frac{\partial \bra{\phi(\vec{\theta}(t))}}{\partial \theta_i}\ket{\phit}\dot\theta_j- \bra{\phit}H\ket{\phi(\vec{\theta}(t))}\right)= C_i^I,
		\end{aligned}
	\end{equation}
	which it is equivalent to
			\begin{equation}
		\begin{aligned}
			&\left(A_{i,j}^R+ \frac{\partial \bra{\phi(\vec{\theta}(t))}}{\partial \theta_i}\ket{\phit}\frac{\partial \bra{\phi(\vec{\theta}(t))}}{\partial \theta_j}\ket {\phit}\right)\dot\theta_j=C_i^I+i\frac{\partial \bra{\phi(\vec{\theta}(t))}}{\partial \theta_i}\ket{\phit}\bra{\phit}H\ket{\phi(\vec{\theta}(t))}.
		\end{aligned}
	\end{equation}
\end{proof}

\section{Imaginary time evolution}
\subsection{Imaginary time evolution: pure states}
\subsubsection{The Dirac and Frenkel variational principle}
Replace real time with imaginary time $\tau = it$, the Wick-rotated Schr\"odinger equation is,
\begin{equation}\label{Schrodinger2}
	\frac{d \ket{\psi(\tau)}}{d \tau} = -(H-E_\tau)\ket{\psi(\tau)},
\end{equation}
where $E_\tau = \braket{{\psi(\tau)}|H|{\psi(\tau)}}$.
Consider a normalised parametrised trial state $\ket{\phitau}$, with $\vec{\theta}(\tau) = (\theta_1(\tau), \theta_2(\tau),\dots, \theta_N(\tau))$, the imaginary time evolution of the Schr\"odinger equation on the trial state space is
\begin{equation}\label{diffEq2}
	\frac{d \ket{\phitau}}{d \tau} = \sum_i \frac{\partial \ket{\phitau}}{\partial \theta_i}\dot{\theta_i}=-(H-E_\tau)\ket{\phitau}.
\end{equation}
Here $\dot \theta_i = \frac{d\theta_i}{d\tau}$.
Apply the projector
\begin{equation}
	P = \sum_i\frac{\partial \ket{\phitau}}{\partial \theta_i}\frac{\partial \bra{\phitau}}{\partial \theta_i},
\end{equation}
we have
\begin{equation}
	\begin{aligned}
		P\sum_j \frac{\partial \ket{\phi(\vec{\theta}(\tau))}}{\partial \theta_j}\dot{\theta}_j&=-P(H-E_\tau)\ket{\phi(\vec{\theta}(t))},\\
		\sum_i\frac{\partial \ket{\phi(\vec{\theta}(\tau))}}{\partial \theta_i}\frac{\partial \bra{\phi(\vec{\theta}(\tau))}}{\partial \theta_i}\sum_j \frac{\partial \ket{\phi(\vec{\theta}(\tau))}}{\partial \theta_j}\dot{\theta}_j&=-\sum_i\frac{\partial \ket{\phi(\vec{\theta}(\tau))}}{\partial \theta_i}\frac{\partial \bra{\phi(\vec{\theta}(\tau))}}{\partial \theta_i}(H-E_\tau)\ket{\phi(\vec{\theta}(\tau))}.
	\end{aligned}
\end{equation}
For each term of $\frac{\partial \ket{\phi(\vec{\theta}(t))}}{\partial \theta_i}$, the coefficient should satisfy
\begin{equation}
\begin{aligned}
	\sum_j \frac{\partial \bra{\phi(\vec{\theta}(\tau))}}{\partial \theta_i}\frac{\partial \ket{\phi(\vec{\theta}(\tau))}}{\partial \theta_j}\dot{\theta}_j&=-\frac{\partial \bra{\phi(\vec{\theta}(\tau))}}{\partial \theta_i}(H-E_\tau)\ket{\phi(\vec{\theta}(\tau))}.
\end{aligned}
	\end{equation}
Define the matrix elements of $A$ and $C$ as
\begin{equation}
	\begin{aligned}
		A_{i,j} &= \frac{\partial \bra{\phi(\vec{\theta}(\tau))}}{\partial \theta_i}\frac{\partial \ket{\phi(\vec{\theta}(\tau))}}{\partial \theta_j},\\
		C_i' &= \frac{\partial \bra{\phi(\vec{\theta}(\tau))}}{\partial \theta_i}(H-E_\tau)\ket{\phi(\vec{\theta}(\tau))},
	\end{aligned}
\end{equation}
the evolution of parameters is simplified to
\begin{equation}
	\sum_j A_{i,j}\dot{\theta}_j = -C_i'.
\end{equation}

The same equation can be obtained by applying the Dirac and Frenkel variational principle 
\begin{equation}
	\left\langle{ \delta\phitau\bigg|\left(\frac{d}{d \tau}+H-E_\tau\right)\bigg|\phi(\vec{\theta}(\tau))}\right\rangle = 0.
\end{equation}
This is can be verified with 
\begin{equation}
	\bra{\delta \phitau} = \sum_i \frac{\partial \bra{\phi(\vec{\theta}(\tau))}}{\partial \theta_i}\delta{\theta_i},
\end{equation}
and Eq.~\eqref{diffEq2}.

\subsubsection{McLachlan's variational principle}
The McLachlan's variational principle \cite{McLachlan}, applied to imaginary time evolution, is given by
 \begin{equation}
 	\delta \|({d}/{d \tau} + H-E_\tau)\ket{\phitau)}\|=0
 \end{equation}
where
\begin{equation}
\begin{aligned}
&\|({d}/{d \tau} + H-E_\tau)\ket{\phitau)}\|^2\\
=&\left(({d}/{d \tau} + H-E_\tau)\ket{\phitau)}\right)^\dag({d}/{d \tau} + H-E_\tau)\ket{\phitau)},\\
	=&\sum_{i,j}\frac{\partial \bra{\phit}}{\partial \theta_i}\frac{\partial \ket{\phit}}{\partial \theta_j}\dot{\theta}_i^* \dot{\theta}_j+ \sum_{i}\frac{\partial \bra{\phit}}{\partial \theta_i}(H-E_\tau)\ket{\phit}\dot{\theta}_i^*\\
		& \sum_{i}\bra{\phit}(H-E_\tau)\frac{\partial \ket{\phit}}{\partial \theta_i}\dot{\theta}_i +\bra{\phit}	(H-E_\tau)^2\ket{\phit}. 
\end{aligned}
\end{equation}

Focusing on $\|({d}/{d \tau} + H-E_\tau)\ket{\phitau)}\|^2$, when $\dot{\theta}_i$ is complex, we have
\begin{equation}
	\begin{aligned}
		&\delta \|({d}/{d \tau} + H-E_\tau)\ket{\phitau)}\|^2\\
		=& \left(\sum_{j}\frac{\partial \bra{\phit}}{\partial \theta_i}\frac{\partial \ket{\phit}}{\partial \theta_j} \dot{\theta}_j+ \frac{\partial \bra{\phit}}{\partial \theta_i}(H-E_\tau)\ket{\phit}\right)\delta\dot{\theta}_i^*,\\
		&+\left(\sum_{j}\frac{\partial \bra{\phit}}{\partial \theta_j}\frac{\partial \ket{\phit}}{\partial \theta_i} \dot{\theta}_j^*+\frac{\partial \bra{\phit}}{\partial \theta_i}(H-E_\tau)\ket{\phit}\right)\delta\dot{\theta}_i.
\end{aligned}
\end{equation}
Then the evolution is 
\begin{equation}
\begin{aligned}
	\sum_j \frac{\partial \bra{\phi(\vec{\theta}(\tau))}}{\partial \theta_i}\frac{\partial \ket{\phi(\vec{\theta}(\tau))}}{\partial \theta_j}\dot{\theta}_j&=-\frac{\partial \bra{\phi(\vec{\theta}(\tau))}}{\partial \theta_i}(H-E_\tau)\ket{\phi(\vec{\theta}(\tau))}.
\end{aligned}
	\end{equation}

When $\dot{\theta}_i$ can only take real values, we have
\begin{equation}
	\begin{aligned}
		&\delta \|({d}/{d \tau} + H-E_\tau)\ket{\phitau)}\|^2\\
		=&\sum_{j}\left(\frac{\partial \bra{\phit}}{\partial \theta_i}\frac{\partial \ket{\phit}}{\partial \theta_j} + \frac{\partial \bra{\phit}}{\partial \theta_j}\frac{\partial \ket{\phit}}{\partial \theta_i}\right) \dot{\theta}_j\delta\theta_i\\
		&+\left(\frac{\partial \bra{\phit}}{\partial \theta_i}(H-E_\tau)\ket{\phit}+\bra{\phit}(H-E_\tau)\frac{\partial \ket{\phit}}{\partial \theta_i}\right)\delta\theta_ix.
\end{aligned}
\end{equation}
The corresponding evolution equation for parameters is
\begin{equation}
	\begin{aligned}
	&\sum_{j}\left(\frac{\partial \bra{\phit}}{\partial \theta_i}\frac{\partial \ket{\phit}}{\partial \theta_j} + \frac{\partial \bra{\phit}}{\partial \theta_j}\frac{\partial \ket{\phit}}{\partial \theta_i}\right) \dot{\theta}_j\\
	=&-\left(\frac{\partial \bra{\phit}}{\partial \theta_i}(H-E_\tau)\ket{\phit}-\bra{\phit}(H-E_\tau)\frac{\partial \ket{\phit}}{\partial \theta_i}\right),
	\end{aligned}
\end{equation}
and it is equivalent to
\begin{equation}
	\sum_j A^R_{i,j}\dot{\theta}_j = -C_i^R,
\end{equation}
where $A^R_{i,j}$ and $C_i^R$ are the real parts of $A_{i,j}$ and $C_i$, respectively. Due to the normalisation requirement for $\phitau$, i.e., $\bra{\phi(\theta(\tau))}\phitau\rangle = 1$, we have $\Re\left(\frac{\partial \bra{\phit}}{\partial \theta_i}E_\tau\ket{\phit}\right) = 0$. Therefore, 
\begin{equation}
	C_i^R=C_i'^R = \Re\left(\frac{\partial \bra{\phit}}{\partial \theta_i}H\ket{\phit}\right).
\end{equation}

\subsubsection{Time-dependent variational principles}
To recover the Wick-rotated Schr\"odinger equation, the corresponding Lagrangian should be modified to be 
\begin{equation}
	L = \left\langle{\psi}\bigg|\frac{d}{d \tau}+H\bigg|\psi\right\rangle+\braket{\psi|H|\psi}(1-\braket{\psi|\psi}).
\end{equation}
Take $\ket{\psi}$, $\frac{d\ket{\psi}}{d\tau}$, $\bra{\psi}$, and $\frac{d\bra{\psi}}{d\tau}$ as free parameters, the Euler-Lagrange equation to $\bra{\psi}$ is
\begin{equation}
	\begin{aligned}
		\frac{\partial L}{\partial \bra{\psi}} - \frac{d}{d\tau}\frac{\partial L}{\partial \frac{\partial\bra{\psi}}{\partial \tau}} &=\left(\frac{d}{d \tau}+H\right)\ket{\psi}+H\ket{\psi}(1-\braket{\psi|\psi})-\braket{\psi|H|\psi}\ket{\psi},\\
		&=\left(\frac{d}{d \tau}+H-E_\tau\right)\ket{\psi},
	\end{aligned}
\end{equation}
where the second line makes use of the normalisation condition $\braket{\psi|\psi}=1$ and the definition of $E_\tau = \braket{\psi|H|\psi}$.

Now, replace $\psi$ with the parameterised trial state $\phitau$, the Lagrangian becomes,
\begin{equation}
\begin{aligned}
	L &= \left\langle{{\phitau}\bigg|\left(\frac{d}{d \tau}+H\right)}\bigg|{\phitau}\right\rangle - \braket{\phitau|H|\phitau}(1-\braket{\phitau|\phitau}),\\
	&=\sum_i\dot{\theta_i}\left\langle{{\phitau}\bigg|\frac{\partial}{\partial \theta_i}}\bigg|{\phitau}\right\rangle - \braket{\phitau|H|\phitau}(2-\braket{\phitau|\phitau})
\end{aligned}
	\end{equation}
Suppose $\theta$ is complex, then 	\begin{equation}
\begin{aligned}
	\frac{\partial L}{\partial \theta_i^*} - \frac{d}{dt}\frac{\partial L}{\partial \dot\theta_i^*} &= \sum_j\dot{\theta_j}\frac{\partial\bra{\phit}}{\partial \theta_i}\frac{\partial\ket{\phit}}{\partial \theta_j}-\frac{\partial \bra{\phit}}{\partial \theta_i}H\ket{\phit}(2-\braket{\phitau|\phitau})\\
	&+\braket{\phitau|H|\phitau}\frac{\partial \bra{\phit}}{\partial \theta_i}\ket{\phit},
\end{aligned} 
\end{equation}
and the evolution is
\begin{equation}
	\sum_j \frac{\partial \bra{\phi(\vec{\theta}(\tau))}}{\partial \theta_i}\frac{\partial \ket{\phi(\vec{\theta}(\tau))}}{\partial \theta_j}\dot{\theta}_j=-\frac{\partial \bra{\phi(\vec{\theta}(\tau))}}{\partial \theta_i}(H-E_\tau)\ket{\phi(\vec{\theta}(\tau))}.
\end{equation}

When 	$\theta_i$ is real, we can have the evolution of parameters with the Euler-Lagrange equation $\frac{\partial L}{\partial \theta_i} - \frac{d}{d\tau}\frac{\partial L}{\partial \dot\theta_i}=0$. The first term $\frac{\partial L}{\partial \theta_i}$
\begin{equation}
\begin{aligned}
	\frac{\partial L}{\partial \theta_i}  =&\sum_j\dot{\theta}_j\frac{\partial }{\partial \theta_i}\left\langle{{\phitau}\bigg|\frac{\partial}{\partial \theta_j}}\bigg|{\phitau}\right\rangle - \frac{\partial }{\partial \theta_i}[\braket{\phitau|H|\phitau}(2-\braket{\phitau|\phitau})],\\
	=&\sum_j\dot{\theta}_j\left(\frac{\partial \bra{\phitau}}{\partial \theta_i}\frac{\partial \ket{\phitau}}{\partial \theta_j}+\bra{\phitau}\frac{\partial^2 \ket{\phitau}}{\partial \theta_i\partial \theta_j}\right)\\
	&-\frac{\partial \bra{\phitau}}{\partial \theta_i}H\ket{\phitau} - \bra{\phitau}H\frac{\partial \ket{\phitau}}{\partial \theta_i}
	\end{aligned} 
\end{equation}
Here, we applied the normalisation condition $\braket{\phitau|\phitau} = 1$ and $\frac{\partial }{\partial \theta_i}(2-\braket{\phitau|\phitau}) = 0$.
The second term $\frac{d}{dt}\frac{\partial L}{\partial \dot\theta_i}$ is
\begin{equation}
\begin{aligned}
	\frac{d}{d\tau}\frac{\partial L}{\partial \dot\theta_i} &=\frac{d}{d\tau}\left\langle{{\phitau}\bigg|\frac{\partial}{\partial \theta_i}}\bigg|{\phitau}\right\rangle,\\
	&=
	\sum_{j} \dot{\theta}_j\left(\frac{\partial\bra{\phitau}}{\partial \theta_j}\frac{\partial\ket{\phitau}}{\partial \theta_i} 
	+\bra{\phitau}\frac{\partial^2 \ket{\phitau}}{\partial \theta_i\partial \theta_j}\right),
	\end{aligned} 
\end{equation}
and
\begin{equation}
\begin{aligned}
	\frac{\partial L}{\partial \theta_i} - \frac{d}{d\tau}\frac{\partial L}{\partial \dot\theta_i}=&
	\sum_{j} \dot{\theta}_j\left(\frac{\partial \bra{\phitau}}{\partial \theta_i}\frac{\partial \ket{\phitau}}{\partial \theta_j}-\frac{\partial\bra{\phitau}}{\partial \theta_j}\frac{\partial\ket{\phitau}}{\partial \theta_i} 
	\right)\\
	&-\frac{\partial \bra{\phitau}}{\partial \theta_i}H\ket{\phitau} - \bra{\phitau}H\frac{\partial \ket{\phitau}}{\partial \theta_i}
	\end{aligned} 
\end{equation}
The evolution of parameters is
\begin{equation}
\begin{aligned}
	&\sum_{j} \dot{\theta}_j\left(\frac{\partial \bra{\phitau}}{\partial \theta_i}\frac{\partial \ket{\phitau}}{\partial \theta_j}-\frac{\partial\bra{\phitau}}{\partial \theta_j}\frac{\partial\ket{\phitau}}{\partial \theta_i} 
	\right)\\
	=&\frac{\partial \bra{\phitau}}{\partial \theta_i}H\ket{\phitau} + \bra{\phitau}H\frac{\partial \ket{\phitau}}{\partial \theta_i},
	\end{aligned} 
\end{equation}
and it is equivalent to
\begin{equation}
		i\sum_j A_{i,j}^I \dot\theta_j = C_i^R. 
\end{equation}
Here, $A^I_{i,j}$ and $C_i^R$ is the imaginary and real parts of $A_{i,j}$ and $C_i$, respectively. Note that the equation always lead to an incorrect imaginary solution of $\dot\theta_j$. 

\subsection{Imaginary time evolution of mixed state}
\subsubsection{Evolution of mixed states}
Under the imaginary time evolution for Hamiltonian $H$, the state $\rho(\tau)$ at imaginary time $\tau$ should be 
\begin{equation}\begin{aligned}
\rho(\tau)= \frac{e^{-H \tau} \rho(0) e^{-H \tau}}{\mathrm{Tr}[e^{-2 H \tau} \rho(0)]},  
\end{aligned} \end{equation}
where $\rho(0)$ is the initial state. It can be easily checked  that $\rho(\tau)$ follows the time derivative equation 
\begin{equation}\begin{aligned}
\frac{ d\rho (\tau)}{d \tau}  = -\bigg( \{H, \rho(\tau)\} - 2\braket{H} \rho(\tau) \bigg), 
\label{imagerho}
\end{aligned} \end{equation}
where $\{H, \rho(\tau)\} = H\rho(\tau)+\rho(\tau)H$.
Consider a parametrised state, and we can obtain the evolution of $\vec{\theta}$ that simulates Eq. (\ref{imagerho}),  
\begin{equation}\begin{aligned}
\sum_i \frac{\partial \rho (\vec{\theta})}{\partial \theta_i} \dot{\theta}_i = -\bigg( \{H, \rhot\} - 2\braket{H} \rhot \bigg) 
\end{aligned} \end{equation}

\paragraph{The Dirac and Frenkel variational principle}
Following the Dirac and Frenkel variational principle, we have 
\begin{equation}\begin{aligned}
&\mathrm{Tr}\bigg[\delta \rho (\vec{\theta}) \bigg( \sum_i \frac{\partial \rho (\vec{\theta})}{\partial \theta_i} \dot{\theta}_i  +\bigg( \{H, \rhot\} - 2\braket{H} \rhot \bigg) \bigg)  \bigg] = 0
\end{aligned} \end{equation}
The evolution of parameters is 
\begin{equation}\begin{aligned}
\sum_j M_{i,j} \dot{\theta}_j = Y_i
\end{aligned} \end{equation}
with 
\begin{equation}\begin{aligned}
M_{i,j}&=\mathrm{Tr} \bigg[ \bigg( \frac{\partial \rho (\vec{\theta})}{\partial \theta_i} \bigg)^\dag \frac{\partial \rhot}{\partial \theta_j} \bigg] \\
Y_i &= -\mathrm{Tr}\bigg[ \bigg( \frac{\partial \rho (\vec{\theta})}{\partial \theta_i} \bigg)^\dag\bigg( \{H, \rhot\} -2 \braket{H} \rhot \bigg) \bigg]
\end{aligned} \end{equation}
Note that when the parameter $\vec{\theta}$ is real, $M$ and $Y$ are also real, and
\begin{equation}
	V_i'=-\mathrm{Tr}\bigg[ \frac{\partial \rhot}{\partial \theta_i}  \{H, \rhot\} \bigg].
\end{equation}
In this case, the solution $\vec{\dot{\theta}}$ is also real.

\paragraph{MacLachlan's variational principle}
We can also minimise the distance between the evolution via parameters and the evolution via Eq. (\ref{imagerho}). 
\begin{equation}\begin{aligned}
\delta \| d \rho/ d \tau +(\{H, \rho\} - 2 \braket{H} \rho) \| =0 
\end{aligned} \end{equation}
where 
\begin{equation}\begin{aligned}
& \delta \| d \rho/d\tau +(\{H, \rho\} - 2 \braket{H} \rho) \|^2 \\
& =\mathrm{Tr} \bigg[ \bigg(d \rho / d \tau +\bigg(\{H, \rho\} -2\braket{H} \rho  \bigg) \bigg)^\dag \bigg(d \rho / d \tau +\bigg(\{H, \rho\} -2\braket{H} \rho  \bigg) \bigg)  \bigg] \\
&= \mathrm{Tr} \bigg[ \bigg( \sum_i \frac{\partial \rho}{\partial \theta_i} \dot{\theta}_i  \bigg)^\dag \sum_j \frac{\partial \rho}{\partial \theta_j} \dot{\theta}_j + \bigg( \sum_i \frac{\partial \rho}{\partial \theta_i} \dot{\theta}_i  \bigg)^\dag \bigg(\{H, \rho\}  - 2\braket{H} \rho \bigg) \bigg] \\ 
&+ \mathrm{Tr}\bigg[ \bigg( \{H, \rho\}  -2 \braket{H} \rho \bigg) \sum_j \frac{\partial \rho}{\partial \theta_j} \dot{\theta}_j  + \bigg( \{H, \rho\}  -2 \braket{H} \rho \bigg)^2  \bigg] 
\end{aligned} \end{equation}
Focusing on $\delta \| d \rho/d\tau +(\{H, \rho\} - 2 \braket{H} \rho) \|^2$, suppose $\dot{\vec{ \theta}}$ is complex, then 
\begin{equation}\begin{aligned}
&\delta \| d \rho/ d \tau +(\{H, \rho\}  - 2 \braket{H} \rho) \|^2  \\
&= \sum_i \bigg( \mathrm{Tr}\bigg[\bigg(\frac{\partial \rho (\vec{\theta})}{\partial \theta_i}\bigg)^\dag \sum_j \frac{\partial \rho (\vec{\theta})}{\partial \theta_j} \dot{\theta}_j +\bigg(\frac{\partial \rho (\vec{\theta})}{\partial \theta_i}\bigg)^\dag \bigg( \{H, \rhot\} -2\braket{H} \rho (\vec{\theta})\bigg) \bigg]   \delta \theta_i^* \\ 
&+ \mathrm{Tr}\bigg[\bigg(\sum_j \frac{\partial \rho (\vec{\theta})}{\partial \theta_j}\dot{\theta}_j \bigg)^\dag \frac{\partial \rho (\vec{\theta})}{\partial \theta_i} +\bigg( \{H, \rhot\} -2\braket{H} \rho (\vec{\theta})\bigg)\frac{\partial \rho (\vec{\theta})}{\partial \theta_i}  \bigg]  \delta \theta_i \bigg)
\end{aligned} \end{equation}
The evolution to $\dot{\vec{\theta}}$ is 
\begin{equation}\begin{aligned}
\sum_j \mathrm{Tr} \bigg[ \bigg( \frac{\partial \rho (\vec{\theta})}{\partial \theta_i} \bigg)^\dag \frac{\partial \rho (\vec{\theta})}{\partial \theta_j} \bigg] \dot{\theta}_j = - \mathrm{Tr} \bigg[\bigg( \frac{\partial \rho (\vec{\theta})}{\partial \theta_i} \bigg)^\dag \bigg(\{H, \rhot\} -2 \braket{H} \rho (\vec{\theta}) \bigg) \bigg]
\end{aligned} \end{equation}
When $\dot{\vec{\theta}}$ is real, then 
\begin{equation}\begin{aligned}
&\delta \| d \rho/d\tau +(\{H, \rho\} - 2 \braket{H} \rho) \|^2 \\
&= \sum_i \bigg( \mathrm{Tr}\bigg[\frac{\partial \rho (\vec{\theta})}{\partial \theta_i} \sum_j \frac{\partial \rho (\vec{\theta})}{\partial \theta_j} \dot{\theta}_j +\frac{\partial \rho (\vec{\theta})}{\partial \theta_i} \bigg( \{H, \rhot\} -2\braket{H} \rho (\vec{\theta})\bigg) \bigg]   \delta \theta_i \\ 
&+ \mathrm{Tr}\bigg[\bigg(\sum_j \frac{\partial \rho (\vec{\theta})}{\partial \theta_j}\dot{\theta}_j \bigg)  \frac{\partial \rho (\vec{\theta})}{\partial \theta_i} +\bigg( \{H, \rhot\} -2\braket{H} \rho (\vec{\theta})\bigg)\frac{\partial \rho (\vec{\theta})}{\partial \theta_i}  \bigg]  \delta \theta_i \bigg)
\end{aligned} \end{equation}
The evolution to $\dot{\vec{\theta}}$ is 
\begin{equation}\begin{aligned}
\sum_j \mathrm{Tr} \bigg[ \frac{\partial \rhot}{\partial \theta_i} \frac{\partial \rhot}{\partial \theta_j} \bigg] \dot{\theta}_j &= -\mathrm{Tr}\bigg[ \frac{\partial \rhot}{\partial \theta_i} \bigg( \{H, \rhot\} -2 \braket{H} \rhot \bigg) \bigg],\\
&=-\mathrm{Tr}\bigg[ \frac{\partial \rhot}{\partial \theta_i}  \{H, \rhot\} \bigg].
\end{aligned} \end{equation}
Then we can obtain the same equation as derived by using the time-dependent variational principle.

\paragraph{The time-dependent variational principle}
The Lagrangian for Eq. (\ref{imagerho}) is 
\begin{equation}\begin{aligned}
L=\mathrm{Tr} \big[\rho ^\dag (d \rho / d \tau +(\{H, \rho\} -2\braket{H} \rho )) \big]
\end{aligned} \end{equation}
Eq. (\ref{imagerho}) is obtained by 
\begin{equation}\begin{aligned}
\frac{\partial L}{\partial \rho^\dag}- \frac{d}{d\tau} \frac{\partial L}{\partial \frac{\rho^\dag}{dt}}= d \rho / d \tau +(\{H, \rho\} -2\braket{H} \rho )=0
\end{aligned} \end{equation}
Suppose $\rho=\rhot$, then the Lagrangian is 
\begin{equation}\begin{aligned}
L=\mathrm{Tr} \bigg[\rhot^\dag \bigg( \bigg(\sum_j \frac{\partial \rhot}{\partial \theta_j} \dot{\theta}_j +\bigg(\{H, \rhot\} -2\braket{H} \rhot \bigg)\bigg) \bigg]
\end{aligned} \end{equation}
Suppose $\vec{\theta}$ is complex, then 
\begin{equation}\begin{aligned}
\frac{\partial L}{\partial \theta_i^*}-\frac{d}{d\tau}\frac{\partial L}{\partial \dot{\theta}_i^*}= \mathrm{Tr}\bigg[\frac{\partial \rhot^\dag}{\partial \theta_i^*}\bigg(\sum_j \frac{\partial \rhot}{\partial \theta_j} \dot{\theta}_j+\bigg(\{H, \rhot\}-2\braket{H}\rhot  \bigg) \bigg) \bigg],
\end{aligned} \end{equation}
and the evolution is
\begin{equation}\begin{aligned}
\sum_j \mathrm{Tr}\bigg[\bigg(\frac{\partial \rhot}{\partial \theta_i} \bigg)^\dag \frac{\partial \rhot}{\partial \theta_j} \bigg] \dot{\theta}_j =-\mathrm{Tr}\bigg[\bigg(\frac{\partial \rhot}{\partial \theta_i} \bigg)^\dag \bigg( \{H, \rhot\}-2\braket{H}\rhot \bigg) \bigg]
\end{aligned} \end{equation}
Instead, suppose $\vec{\theta}$ is real, we cannot reproduce the imaginary time evolution as we have $\rho=\rho^\dag$. Therefore, we cannot  the equation for evaluating $\dot{\vec{\theta}}$.

\subsubsection{Reduction to the pure state case}
Suppose $\rhot$ is a pure state $\ket{\phi(\vec{\theta})}\bra{\phi(\vec{\theta})}$, and Eq. (\ref{imagerho}) should become the Wicked-rotated Schr\"odinger equation. For the variational evolution equation, we have
\begin{equation}
	\begin{aligned}
		M_{i,j} &= \tr\left[\frac{\partial \rhot}{\partial \theta_i}\frac{\partial \rhot}{\partial \theta_j}\right],\\
		 &= \tr\left[\left(\frac{\partial \ket{\phi(\vec{\theta}(\tau))}}{\partial \theta_i}\bra{\phi(\vec{\theta}(\tau))}+\ket{\phi(\vec{\theta}(\tau))}\frac{\partial \bra{\phi(\vec{\theta}(\tau))}}{\partial \theta_i}\right)\left(\frac{\partial \ket{\phi(\vec{\theta}(\tau))}}{\partial \theta_j}\bra{\phi(\vec{\theta}(\tau))}+\ket{\phi(\vec{\theta}(\tau))}\frac{\partial \bra{\phi(\vec{\theta}(\tau))}}{\partial \theta_j}\right)\right],\\
		&= \Re\left(\frac{\partial \bra{\phi(\vec{\theta}(\tau))}}{\partial \theta_i}\frac{\partial \ket{\phi(\vec{\theta}(\tau))}}{\partial \theta_j}+\frac{\partial \bra{\phi(\vec{\theta}(\tau))}}{\partial \theta_i}\ket{\phi(\vec{\theta}(\tau))}\frac{\partial \bra{\phi(\vec{\theta}(\tau))}}{\partial \theta_j}\ket {\phi(\vec{\theta}(\tau))}\right) ,\\ 
		&= A_{i,j}^R+ \frac{\partial \bra{\phi(\vec{\theta}(\tau))}}{\partial \theta_i}\ket{\phi(\vec{\theta}(\tau))}\frac{\partial \bra{\phi(\vec{\theta}(\tau))}}{\partial \theta_j}\ket {\phi(\vec{\theta}(\tau))},\\
		V_i^\prime &= -\tr\left[\frac{\partial \rhot}{\partial \theta_i}\bigg(\{H, \rhot\}\bigg)\right],\\
		&= -\tr\left[\left(\frac{\partial \ket{\phi (\vec{\theta}(\tau))}}{\partial \theta_i}\bra{\phi(\vec{\theta}(\tau))} +\ket{\phi(\vec{\theta}(\tau))}\frac{\partial \bra{\phi(\vec{\theta}(\tau))}}{\partial \theta_i}\right)\left(H\ket{\phi(\vec{\theta}(\tau))}\bra{\phi(\vec{\theta}(\tau))} + \ket{\phi(\vec{\theta}(\tau))}\bra{\phi(\vec{\theta}(\tau))} H \right)\right],\\
		&= - C_i ^R
	\end{aligned} 
\end{equation}
This equation is equivalent to applying a phase gate to an additional ancilla with an additional parameter in the pure state equation.
\begin{proof}
	Suppose there are $N$ parameters in the ansatz. Consider an additional ancilla that is rotated via a phase gate $e^{i\theta_{0}}$, the evolution of the $N+1$ parameters via the pure state equation is equivalent to the evolution of the $N$ parameters via the mixed state equation. Under the pure state equation, the evolution of the $N+1$ parameters is
	\begin{equation}
		\begin{aligned}
			\sum_{j=1}^N A^R_{i,j}\dot\theta_j + i\frac{\partial \bra{\phi(\vec{\theta}(\tau))}}{\partial \theta_i}\ket{\phi(\vec{\theta}(\tau))}\dot\theta_{0}&=- C_i^R,\\
			-i\sum_{j=1}^N \frac{\partial \bra{\phi(\vec{\theta}(\tau))}}{\partial \theta_i}\ket{\phi(\vec{\theta}(\tau))}\dot\theta_j-\dot\theta_{0}&= 0.
		\end{aligned}
	\end{equation}
	Substituting the last equation into the first $N$ equations, we get
		\begin{equation}
		\begin{aligned}
			&\sum_{j=1}^N A^R_{i,j}\dot\theta_j + i\frac{\partial \bra{\phi(\vec{\theta}(\tau))}}{\partial \theta_i}\ket{\phi(\vec{\theta}(\tau))}\left(-i\sum_{j=1}^N \frac{\partial \bra{\phi(\vec{\theta}(\tau))}}{\partial \theta_i}\ket{\phi(\vec{\theta}(\tau))}\dot\theta_j\right)=- C_i^R,
		\end{aligned}
	\end{equation}
	which it is equivalent to
			\begin{equation}
		\begin{aligned}
			&\left(A_{i,j}^R+ \frac{\partial \bra{\phi(\vec{\theta}(\tau))}}{\partial \theta_i}\ket{\phi(\vec{\theta}(\tau))}\frac{\partial \bra{\phi(\vec{\theta}(\tau))}}{\partial \theta_j}\ket {\phi(\vec{\theta}(\tau))}\right)\dot\theta_j=-C_i^R.
		\end{aligned}
	\end{equation}
        \end{proof}

\bibliographystyle{unsrtnat}
\bibliography{VariationalTheory}

\end{document}